\documentclass[%
 aip,
amsmath,amssymb,
 reprint,%
]{revtex4-1}

\usepackage{graphicx}
\usepackage{dcolumn}
\usepackage{bm}

\usepackage[utf8]{inputenc}
\usepackage[T1]{fontenc}
\usepackage{mathptmx}
\usepackage{etoolbox}
\usepackage{xcolor}

\makeatletter
\def\@email#1#2{%
 \endgroup
 \patchcmd{\titleblock@produce}
  {\frontmatter@RRAPformat}
  {\frontmatter@RRAPformat{\produce@RRAP{*#1\href{mailto:#2}{#2}}}\frontmatter@RRAPformat}
  {}{}
}%
\makeatother
\begin{document}

\preprint{AIP/123-QED}

\title{h-CMD: An efficient hybrid fast centroid and quasi-centroid molecular dynamics method for the simulation of vibrational spectra}

\author{Dil K. Limbu$^{\dagger}$}
\author{Nathan London$^{\dagger}$}
\author{Md Omar Faruque}
\author{Mohammad R. Momeni}
\email{mmomenitaheri@umkc.edu}
\date{}
\affiliation{Division of Energy, Matter and Systems, School of Science and Engineering, University of Missouri $-$ Kansas City, Kansas City 64110, MO United States.}

\date{\today}

\begin{abstract}
Developing efficient path integral (PI) methods for atomistic simulations of vibrational spectra in heterogeneous condensed phases and interfaces has long been a challenging task. Here, we present the h-CMD method, short for hybrid centroid molecular dynamics, that combines the recently introduced fast quasi-CMD (f-QCMD) method with fast CMD (f-CMD). In this scheme, molecules that are believed to suffer more seriously from the curvature problem of CMD, e.g., water, are treated with f-QCMD, while the rest, e.g., solid surfaces, are treated with f-CMD. To test the accuracy of the newly introduced scheme, the infrared spectra of the interfacial D$_2$O confined in the archetypal ZIF-90 framework are simulated using h-CMD compared to a variety of other PI methods, including thermostatted ring-polymer molecular dynamics (T-RPMD) and partially adiabatic CMD as well as f-CMD and experiment as reference. Comparisons are also made to classical MD, where nuclear quantum effects are neglected entirely. Our detailed comparisons at different temperatures of 250-600 K show that h-CMD produces O--D stretches that are in close agreement with the experiment, correcting the known curvature problem and red-shifting of the stretch peaks of CMD. h-CMD also corrects the known issues associated with too artificially dampened and broadened spectra of T-RPMD, which leads to missing the characteristic doublet feature of the interfacial confined water, rendering it unsuitable for these systems. The new h-CMD method broadens the applicability of f-QCMD to heterogeneous condensed phases and interfaces, where defining curvilinear coordinates for the entire system is not feasible.
\end{abstract}
\maketitle

\section{\label{sec1:intro}Introduction}

Liquid-air, liquid-liquid, and liquid-solid interfaces play critical roles in various chemical, biological, materials, and environmental processes. Vibrational infrared (IR) spectroscopy provides a wealth of information about interactions of molecules in condensed phases and at different interfaces, especially in aqueous environments.\cite{nl_8_2959, pccp_13_17658, pccp_16_12202, cr_116_7590} Specifically, IR spectra can reveal key fundamental insights into the complex hydrogen bond (HB) network of water and how it is influenced by temperature, pH, reducing or oxidizing potentials, and the presence of different ions, among others. Reliable atomistic molecular dynamics (MD) simulation of vibrational spectra requires an efficient way of incorporating nuclear quantum effects (NQEs). However, while the role of NQEs in chemical and material sciences is a widely acknowledged fact, the development of efficient computational methods and software platforms capable of incorporating them into MD simulations of complex systems has been a challenging task.\cite{nrc_2_0109} Path-integral approaches, including path-integral MD (PIMD)\cite{janParrinello1984}, centroid MD (CMD),\cite{cmd1, cmd2, cmd3, cmd4} and ring polymer MD (RPMD),\cite{Craig:2004, Craig:2005a, Craig:2005b} as well as a myriad of more sophisticated methods branched from them,\cite{arpc_64_387} have emerged as the leading platform for the incorporation of NQEs in large-scale condensed-phase simulations. In these methods, NQEs are captured using Feynman’s imaginary-time path-integral formalism~\cite{feynman:1965,feynman:1972} by taking advantage of the so-called isomorphism between the quantum-mechanical partition function and the partition function of a classical ring polymer (RP). Every nuclear degree of freedom (DOF) is represented by an RP comprised of $n$ copies of the system, known as “beads”, connected by harmonic springs. The extended phase space of this RP captures NQEs, such as zero-point energy and nuclear tunneling. Hence, accurate yet computationally efficient path integral (PI) techniques provide powerful alternatives to prohibitively expensive full quantum-mechanical dynamical simulations.

However, when it comes to calculating the IR spectra of complex heterogeneous condensed phases and interfaces, the quantum dynamic community has struggled to propose a computationally efficient PI method that can accurately capture NQEs, which are more pronounced in spectra than other structural and dynamical properties. One of the most well-known formalisms to approximate the real-time dynamics while incorporating NQEs in the PI framework is CMD. In CMD, the dynamics of the particles are performed under the effective mean-field potential of the imaginary-time path-integral whose centroid is constrained to be at the position of the particle. For low dimensional systems, this effective potential can be calculated on a grid prior to dynamical simulations, while for large systems, it must be calculated “on-the-fly,” which is referred to as adiabatic CMD (ACMD).\cite{cmd3} This is achieved by introducing an adiabatic separation between the centroid and internal modes by using the physical mass of the particle as the centroid mass, scaling down the mass of the internal modes of the RP, and attaching them to a thermostat so that they sample the equilibrium distribution while being constrained to the position of the slower moving centroid.\cite{cmd3} Full adiabatic separation in CMD requires a very small mass for the internal modes and, thus, a very small simulation time step for the dynamics to be accurate. The computational cost associated with such choices led to the development of the partially adiabatic CMD (PA-CMD) method, where the mass scaling is not as extreme, allowing for a larger time step while still providing accurate dynamics.\cite{jcp_124_154103} Further work in reducing the computational cost of CMD comes in the form of fast CMD (f-CMD), where the effective potential is learned beforehand from PIMD simulations using force matching to create a new classical analytical force field~\cite{janHone2005} or neural network potential (NNP)~\cite{octLoose2022}. 

It has been documented that vibrational spectra calculated using CMD methods exhibit red-shifting of peaks, primarily of stretch bands, which is worsened with the lowering of temperatures and/or for lighter particles.\cite{Marx_2010} This phenomenon, dubbed the “curvature problem”, is a result of the RP spreading out along the angular coordinates of a system when approaching the inside of the stretching coordinates. This stretching reduces the force along the stretching coordinate, lowering the oscillation frequency of that coordinate and red-shifting the spectra. One approach to overcome the curvature problem is the quasi-centroid MD (QCMD) method introduced by Althorpe and coworkers.~\cite{qcmd,qcmd_nh3,qcmd_torque} QCMD works by defining a quasi-centroid (QC) for the molecule using curvilinear coordinates and determining the effective mean-field potential using an imaginary time RP that has its QC constrained to the particle position as opposed to its Cartesian centroid.~\cite{qcmd} The resulting potential does not exhibit the same flattening on the inside of stretching curves as in CMD and thus does not have the red-shifting of vibrational spectra. The original adiabatic implementation of QCMD does not lend itself to scaling to complex chemical systems due to computational complexity and cost.~\cite{qcmd,qcmd_torque} Inspired by the f-CMD approach, Manolopoulos and coworkers developed the f-QCMD method.~\cite{decFletcher2021,fqcmd_water_ice} In f-QCMD, a corrective potential to the classical force field is learned to create QCMD-like dynamics in classical simulations, but the corrections are found using iterative Boltzmann inversion (IBI)~\cite{janSoper1996,Reith2003} on a set of distribution functions instead of force matching. While f-QCMD has shown reduced computational cost compared to QCMD,~\cite{decFletcher2021,fqcmd_water_ice} determining the distribution functions for large molecules/frameworks poses a challenge. In a similar vein, Kapil and coworkers developed the temperature elevation PI coarse-graining simulations (T$_e$ PIGS) method.~\cite{tepigs} Like with some f-CMD implementations, a classical NNP is learned through force matching a PIMD simulation to incorporate NQEs. However, the PIMD simulations are performed at a high temperature where the RP distribution is more compact.~\cite{tepigs} The resulting NNP is then used for calculating vibrational spectra at lower temperatures without the curvature problem of CMD.~\cite{tepigs} While offering a simpler implementation for complex systems compared to f-QCMD, performing the high-temperature simulations to obtain the NNP could pose challenges for delicate systems or when using an NNP for the PIMD simulation where bond breaking/forming is possible at high temperatures.

Here, to circumvent the curvature problem and create an accurate yet efficient scheme for calculating vibrational spectra of complex condensed phases and interfaces, we have devised a new method coined hybrid CMD (h-CMD) that combines f-QCMD and f-CMD methods. This method allows us to simulate interfacial systems involving complex frameworks with high accuracy for the most important degrees of freedom and reduced complexity for the others. To bring f-QCMD and f-CMD methods on equal computational footing, the latter f-CMD method is implemented with the regularized IBI method used in f-QCMD. 
We showcase the accuracy of our new hybrid h-CMD method on simulating IR spectra of D$_2$O confined in zeolitic-imidazolate framework-90 (ZIF-90) compared to the experiment as reference. By comparing the spectra using our hybrid h-CMD method to PA-CMD and T-RPMD, as well as f-CMD and classical MD at different temperatures, we provide detailed insights into the applicability of this new scheme to complex interfaces.

This work is structured as follows: In Section~\ref{sec:methods}, we outline our h-CMD method, followed by the methods used for calculating vibrational spectra. Simulation details are provided in Section~\ref{sec:sim-det}. In section~\ref{sec:results}, we provide our detailed analyses of the IR spectra of interfacial D$_2$O confined in ZIF-90 at different temperatures, followed by conclusions and future works.

\section{\label{sec:methods}Methods}

\subsection{\label{sec4:hcmd}Hybridizing f-QCMD and f-CMD methods: h-CMD}

The CMD framework, particularly PA-CMD, stands as a popular method for incorporating NQEs in calculating vibrational spectra despite its known issue of the curvature problem. QCMD has been shown to remedy the curvature problem for some simple chemical systems,~\cite{qcmd,qcmd_nh3,qcmd_torque}, but its computational complexity hinders its applicability for general use. While f-QCMD does a lot of work to simplify the QCMD method,~\cite{decFletcher2021,fqcmd_water_ice} it has its own limitations for its use for complex systems. Here, we introduce a new method that we refer to as hybrid CMD (h-CMD). It combines the f-QCMD and f-CMD methods to create a scheme that is applicable to heterogeneous systems containing large complex molecules and/or extended frameworks. 
In this section, we give an overview of the QCMD and f-QCMD methods and discuss their limitations for complex molecules and extended frameworks. We then give our rationale for the h-CMD method and discuss its formulation.

The original QCMD method overcomes the curvature problem by altering how the effective mean-field potential is defined through the use of curvilinear coordinates for the centroid.~\cite{qcmd} 
To illustrate this, we will discuss these methods in the context of a system of D$_2$O molecules. Each water molecule can be defined using the set of Cartesian coordinates for each of the three atoms, $\{\mathbf{q}_\mathrm{O},\mathbf{q}_{\mathrm{D}_1},\mathbf{q}_{\mathrm{D}_2}\}$. Additionally, the OD bonds and DOD angle are defined as~\cite{qcmd}
  \begin{equation}
    \begin{split}
      r_i &= \left| \mathbf{q}_{\mathrm{OD}_i} \right|,\quad  i=1,2 \\
      \theta &= \mathrm{cos}^{-1}\left( 
      \frac{\mathbf{q}_{\mathrm{OD}_1}\cdot\mathbf{q}_{\mathrm{OD}_2}}{r_1 r_2} \right),
    \end{split}
  \end{equation}
  where $\mathbf{q}_{\mathrm{OD}_i} = \mathbf{q}_{\mathrm{O}} - \mathbf{q}_{\mathrm{D}_i}$.
  These bond and angle values are used to define a set of curvilinear coordinates,
    $\{R_1,R_2,\Theta\}$, that are obtained from bead-averaging
    as,~\cite{qcmd,fqcmd_water_ice}
  \begin{equation}
    R_i = \frac{1}{n}\sum^{n}_{\alpha=1} r_{i,\alpha},
    \label{eq:radCon}
  \end{equation}
  \begin{equation}
    \Theta = \frac{1}{n} \sum^{n}_{\alpha=1} \theta_{\alpha},
    \label{eq:thetaCon}
  \end{equation}
  where $r_{i,\alpha}$ is the $i$-th OD bond using the $\alpha$-th bead, and $\theta_\alpha$ is the
    DOD angle using the $\alpha$-th bead.

The relationship between the Cartesian and polar bead coordinates is non-linear. As such, the set of centroids defined by the polar coordinates are called the quasi-centroids (QCs),
      $\left\{ 
        \overline{\mathbf{Q}}_{\mathrm{O}},
        \overline{\mathbf{Q}}_{\mathrm{D}_1},
        \overline{\mathbf{Q}}_{\mathrm{D}_2}
      \right\}$, 
      and are not exactly equivalent to those of the Cartesian centroids,\cite{augTrenins2019}
      $\left\{ 
        {\mathbf{Q}}_{\mathrm{O}},
        {\mathbf{Q}}_{\mathrm{D}_1},
        {\mathbf{Q}}_{\mathrm{D}_2}
      \right\}$,
      defined as
      \begin{equation}
        \mathbf{Q}_\mathrm{X} = \frac{1}{n}\sum^{n}_{\alpha=1}\mathbf{q}_\mathrm{X},
      \end{equation}
      with $\mathrm{X}\in\{\mathrm{O},\mathrm{D}_1,\mathrm{D}_2\}$.
However, as the RP distribution becomes more compact, the two sets become closer in value and are exactly equal in the classical limit.\cite{augTrenins2019}

For gas phase systems, the forces on the QCs can be found directly through their relationship with the curvilinear coordinates.~\cite{qcmd} Simulations used an adiabatic implementation (AQCMD) where two systems were evolved simultaneously, one for the particles of interest and one of the corresponding RPs, which had their QCs constrained to the positions of the particles.~\cite{qcmd,qcmd_nh3,qcmd_torque} In condensed phase systems, multiple molecules interact with each other, so the positions of the QCs of each molecule relative to each other must be known. While the set of curvilinear coordinates is enough to define the positions of the atoms in a molecule with respect to each other, it cannot uniquely define the molecule's location and orientation in 3D space. QCMD uses an Eckart-like frame~\cite{aprEckart1935,marWilson1980} to orient the molecules by rotating the QCs into the frame of the Cartesian centroid.~\cite{qcmd} These Eckart-like conditions are met through the use of SHAKE and RATTLE algorithms,\cite{marRyckaert1977,octAndersen1983,Tuckerman:2010} while the forces between molecules are introduced through rotational forces.~\cite{qcmd} Approximations to the torque on the QCs are needed,~\cite{qcmd} with work to improve the estimation of the torque being done.~\cite{qcmd_torque}

The AQCMD method suffers from a high computational expense associated with its need to use a small mass and, therefore, a small time step to achieve the adiabatic separation for properly sampling the effective mean-field potential. In an effort to make QCMD more feasible for large-scale simulations, the f-QCMD method has been developed.~\cite{decFletcher2021,fqcmd_water_ice} f-QCMD is inspired by the f-CMD method, where the effective potential is learned beforehand, but the potential is meant to mimic that of QCMD. This allows for classical-like simulations with the same dynamics as those of AQCMD.

The effective potential of f-QCMD has the form,\cite{decFletcher2021,fqcmd_water_ice}
  \begin{equation}
    \label{eq:qc-pot}
    V_{\mathrm{qc}}(\mathbf{r}) = V_{\mathrm{cl}}(\mathbf{r}) + \Delta V_{\mathrm{intra}}(\mathbf{r}) +
      \Delta V_{\mathrm{inter}}(\mathbf{r}),
  \end{equation}

\noindent where $V_{\mathrm{cl}}(\mathbf{r})$ is the base, classical potential, and $\Delta V_{\mathrm{intra}}(\mathbf{r})$ and $\Delta V_{\mathrm{inter}}(\mathbf{r})$ are the correction terms for intra- and inter-molecular interactions, respectively. The intra-molecular correction, for a system of $N$ water molecules, is approximated as corrections in terms of curvilinear coordinates,\cite{decFletcher2021,fqcmd_water_ice}
  \begin{equation}
    \Delta V_{\mathrm{intra}}(\mathbf{r}) \simeq \sum^{N}_{I=1} \sum^{2}_{i=1} \Delta V_R(R_{Ii}) +
      \sum^{N}_{I=1} \Delta V_\Theta(\Theta_I),
    \label{eq:intra-correct}
  \end{equation}
and the inter-molecular correction is taken as a sum of pairwise contributions,\cite{decFletcher2021,fqcmd_water_ice}
  \begin{equation}
    \Delta V_{\mathrm{inter}}(\mathbf{r}) \simeq \sum^{N}_{I=1} \sum^{N}_{J\neq I} \sum_{\mathrm{X} \in I} 
    \sum_{\mathrm{Y}\in J}
    \Delta V_\mathrm{XY}(|\overline{\mathbf{Q}}_{\mathrm{X}I} - \overline{\mathbf{Q}}_{\mathrm{Y}J}|),
    \label{eq:inter-correct}
  \end{equation}
where the sums over $I$ and $J$ are over molecules, with $J\neq I$, and the last two sums are over the atoms of type X and Y (O and H for water) in molecules $I$ and $J$, respectively.

Instead of using force matching as in the established f-CMD method, the correction potentials in f-QCMD are determined using the IBI method~\cite{janSoper1996,Reith2003} with a set of distribution functions of two types.\cite{decFletcher2021,fqcmd_water_ice} The first type is the radial distribution functions of the system, defined as
  \begin{equation}
    \begin{split}
      g_{\mathrm{OO}}(r) &= \frac{V}{2\pi r^2 N^2} 
      \sum_{I=1}^{N} \sum_{J\neq I}^{N}
      \left\langle
      \delta(r-|\overline{\mathbf{Q}}_{\mathrm{O}I}-\overline{\mathbf{Q}}_{\mathrm{O}J}|)\right\rangle \\ 
      g_{\mathrm{OD}}(r) &= \frac{V}{8\pi r^2 N^2} 
      \sum_{I=1}^{N} \sum_{J\neq I}^{N}\sum_{j=1}^2
      \left\langle
      \delta(r-|\overline{\mathbf{Q}}_{\mathrm{O}I}-\overline{\mathbf{Q}}_{\mathrm{D}_jJ}|)\right\rangle \\ 
      g_{\mathrm{DD}}(r) &= \frac{V}{8\pi r^2 N^2} 
      \sum_{I=1}^{N} \sum_{J\neq I}^{N}\sum_{i,j=1}^2
      \left\langle
      \delta(r-|\overline{\mathbf{Q}}_{\mathrm{H}_iI}-\overline{\mathbf{Q}}_{\mathrm{D}_jJ}|)\right\rangle, 
    \end{split}
  \end{equation}
where $V$ is the volume of the simulation cell.

The other type of distribution function needed to compute the correction potentials is the intra-molecular distributions,\cite{decFletcher2021,fqcmd_water_ice}
  \begin{equation}
    \rho_R(r) = \frac{1}{2N}\sum_{I=1}^N\sum_{i=1}^2 \left\langle \delta(r-R_{Ii}) \right\rangle
  \end{equation}
and
  \begin{equation}
    \rho_\Theta(\theta) = \frac{1}{N}\sum_{I=1}^N\left\langle \delta(\theta-\Theta_I) \right\rangle,
  \end{equation}
  where $R_{Ii}$ and $\Theta_I$ are the values from Eqns.~\ref{eq:radCon} and~\ref{eq:thetaCon} for molecule $I$, respectively.

The IBI process contains a series of steps to generate the corrective potentials. First, the distributions described above are obtained from PIMD simulations as a reference to be replicated by the effective potential. From there, an iteration zero, where $\Delta V_{\mathrm{intra}}^{(0)}(\mathbf{r})$ and $\Delta V_{\mathrm{inter}}^{(0)}(\mathbf{r})$ are taken to be zero is performed. This is equivalent to classical dynamics under the potential, $V_{\mathrm{cl}}\mathbf(r)$.  For the $i$-th iteration, the distribution functions, $\rho_{R}^{(i)}(r)$, $\rho_{\Theta}^{(i)}(\theta)$, and $g_{\mathrm{XY}}^{(i)}(r)$ are calculated as classical averages under the effective potential. The potentials for the next iteration are found using~\cite{fqcmd_water_ice} 
  \begin{equation}
    \begin{split}
      \Delta V_R^{(i+1)}(r) &= \Delta V_R^{(i)}(r) - \frac{1}{\beta}\mathrm{ln}\left( 
        \frac{\rho_R^\mathrm{exact}(r)}{\rho_R^{(i)}(r)}\right)\\
        \Delta V_\Theta^{(i+1)}(\theta) &= \Delta V_\Theta^{(i)}(\theta) - \frac{1}{\beta}\mathrm{ln}\left( 
        \frac{\rho_\Theta^\mathrm{exact}(\theta)}{\rho_\Theta^{(i)}(\theta)}\right)\\
          \Delta V_\mathrm{XY}^{(i+1)}(r) &= \Delta V_\mathrm{XY}^{(i)}(r) - \frac{1}{\beta}\mathrm{ln}\left( 
          \frac{g_\mathrm{XY}^\mathrm{exact}(r)}{g_\mathrm{XY}^{(i)}(r)}\right),
    \end{split}
  \end{equation}
where $\beta=1/k_\mathrm{B}T$ and the exact distributions are taken to be those from the PIMD simulations calculated through binning histograms. It is important to stress that the reference calculations are standard PIMD simulations, and the QCs are only calculated to obtain the distribution functions and are not involved in the dynamics. To reduce the computational expense of the IBI process, the radial distribution functions of the classical simulations with the effective potentials are calculated using force sampling\cite{octRotenberg2020} to allow for shorter simulations. During all classical-like simulations, the values of the QCs are taken to be positions of the classical atoms.

The form of the intra- and inter-molecular correction potentials do differ. The intra-molecular corrections are taken to be of the same functional form as the classical potentials.\cite{decFletcher2021,fqcmd_water_ice} For the q-TIP4P/F water potential, the OD bond is a quasi-Morse potential, and the DOD angle is harmonic.~\cite{Habershon:2009} This allows for the parameters of the potential to be updated to reflect the correction. The inter-molecular corrections are treated on a grid of $r$ values that are interpolated during the simulation and added to the standard non-bonded interactions. Additionally, because the IBI iterations can be dominated by statistical errors when the values of the RDFs are small, and the IBI process can become unstable near convergence~\cite{fqcmd_water_ice}, the regularized form of IBI is used for the inter-molecular corrections. The updates to the corrections potentials are now
  \begin{equation}
    \Delta V^{(i+1)}_{\mathrm{XY}}(r) = \Delta V^{(i)}_{\mathrm{XY}}(r) - \frac{1}{\beta}
    \ln\left(\frac{g_{\mathrm{XY}}^{\mathrm{exact}}(r)+\varepsilon G_{\mathrm{XY}}}
    {g_{\mathrm{XY}}^{(i)}(r) + \varepsilon G_{\mathrm{XY}}}\right),
  \end{equation}
where $\varepsilon$ is a positive scalar parameter and $G_\mathrm{XY}$ is the maximum value of the two distribution functions, allowing for the same value of $\varepsilon$ for all RDFs. Larger values of the regularization parameter result in a smaller magnitude in the update to the correction potential, helping to remove instabilities in the IBI procedure, particularly in areas with low density.~\cite{fqcmd_water_ice}
Once the distribution functions of the PIMD reference and the classical simulations under the corrected potential converge, the potential can then be used for future simulations. The IBI process only needs to be performed once for a given system at a given temperature, allowing for computationally inexpensive simulations with accurate NQEs.

In f-QCMD, the Eckart-like conditions are met by calculating a rotation matrix for each molecule.~\cite{decFletcher2021,fqcmd_water_ice} The rotation matrix is chosen to minimize the sum of the mass-weighted squared deviations,
  \cite{augTrenins2019,novTrenins2022,octLawrence2023}
  \begin{equation}
    w(\mathbf{U}) = \sum_\mathrm{X} m_\mathrm{X}
    \left[ \left(\mathbf{Q}_\mathrm{X} - \mathbf{Q}_{\mathrm{COM}}\right)
      - \mathbf{U}\left(\overline{\mathbf{Q}}_\mathrm{X} - \overline{\mathbf{Q}}_{\mathrm{COM}}\right)
    \right]^2,
  \end{equation}
  where the sum runs over all the atoms in the molecule, $m_\mathrm{X}$ is the mass of atom X, and $\mathbf{Q}_{\mathrm{COM}}$ and $\overline{\mathbf{Q}}_{\mathrm{COM}}$ is the center-of-mass (COM) of the molecule using the Cartesian centroid and QC coordinates, respectively. The transformed Cartesian QC coordinates are then found using
  \begin{equation}
    \overline{\mathbf{Q}}^*_\mathrm{X} = 
    \mathbf{Q}_{\mathrm{COM}}
    + \mathbf{U}\left(\overline{\mathbf{Q}}_\mathrm{X} - \overline{\mathbf{Q}}_{\mathrm{COM}}\right).
  \end{equation}
The resulting QC coordinates are positioned such that the molecule's COM is exactly the same as that defined by the
Cartesian centroids and oriented to minimize the difference between $\overline{\mathbf{Q}}_\mathrm{X}$ and $\mathbf{Q}_
\mathrm{X}$, while being constrained to the curvilinear coordinates. The rotation matrix $\mathbf{U}$ can be found in a straightforward manner using quaternion algebra.\cite{aprKrasnoshchekov2014,octLawrence2023}
The primary challenge of applying f-QCMD to a system with large molecules/frameworks is defining the set of initial Cartesian coordinates needed to generate the rotation matrices. 

These limitations of QCMD and f-QCMD have motivated us to create a method that can generalize to complex systems in a straightforward manner. To do so, we note that the curvature problem presents itself the most in stretching modes involving light atoms, like hydrogen, so those modes are the ones that need an accurate treatment the most. For other modes in the system, especially those involving only heavy atoms, the RP distribution is more compact, so the QC and Cartesian centroid are very close to each other. Thus, the relevant QCMD and CMD effective potentials are also very similar to each other. To take advantage of this, we have designed h-CMD so that molecules that require a QCMD-level potential are treated with f-QCMD, while other molecules are treated at a CMD level using f-CMD.
To allow the simultaneous use of f-CMD and f-QCMD, we implemented a new version of f-CMD using IBI to create the effective potential. The methodology of this f-CMD is the same as f-QCMD, but the distribution functions are calculated using the Cartesian centroids instead of the QCs. For the intra-molecular distributions, the bond and  angle values are taken to be the polar coordinates from the Cartesian centroids instead of bead-averaged,
  \begin{equation}
    \begin{split}
      R_i &= \left| \mathbf{Q}_{\mathrm{OD}_i} \right|,\quad  i=1,2 \\
      \Theta &= \mathrm{cos}^{-1}\left( 
      \frac{\mathbf{Q}_{\mathrm{OD}_1}\cdot\mathbf{Q}_{\mathrm{OD}_2}}{R_1 R_2} \right),
    \end{split}
    \label{eq:fCMDpolar}
  \end{equation}
where $\mathbf{Q}_{\mathrm{OD}_i} = \mathbf{Q}_{\mathrm{O}} - \mathbf{Q}_{\mathrm{D}_i}$.
The IBI process can be then carried out to generate the corrective potentials consistent with the CMD effective mean-field potential, allowing for classical-like simulations with CMD NQEs incorporated.

Using this implementation of f-CMD for complex molecules reduces the complexity of calculating the reference distributions, as no rotation matrices will be needed for those molecules. As mentioned above, for heterogeneous systems, treating small molecules where the accuracy of the incorporation of NQEs is important to the spectra with f-QCMD and larger molecules and frameworks with f-CMD is the basis of our h-CMD method. It should be emphasized that the difference in the treatment of the molecules is in how the reference distributions are calculated; all molecules are treated the same in the actual PIMD simulations, the IBI process, and the final classical-like dynamics under the correction potentials.

\begin{figure}[t]
\centering
\includegraphics[width=0.99\linewidth]{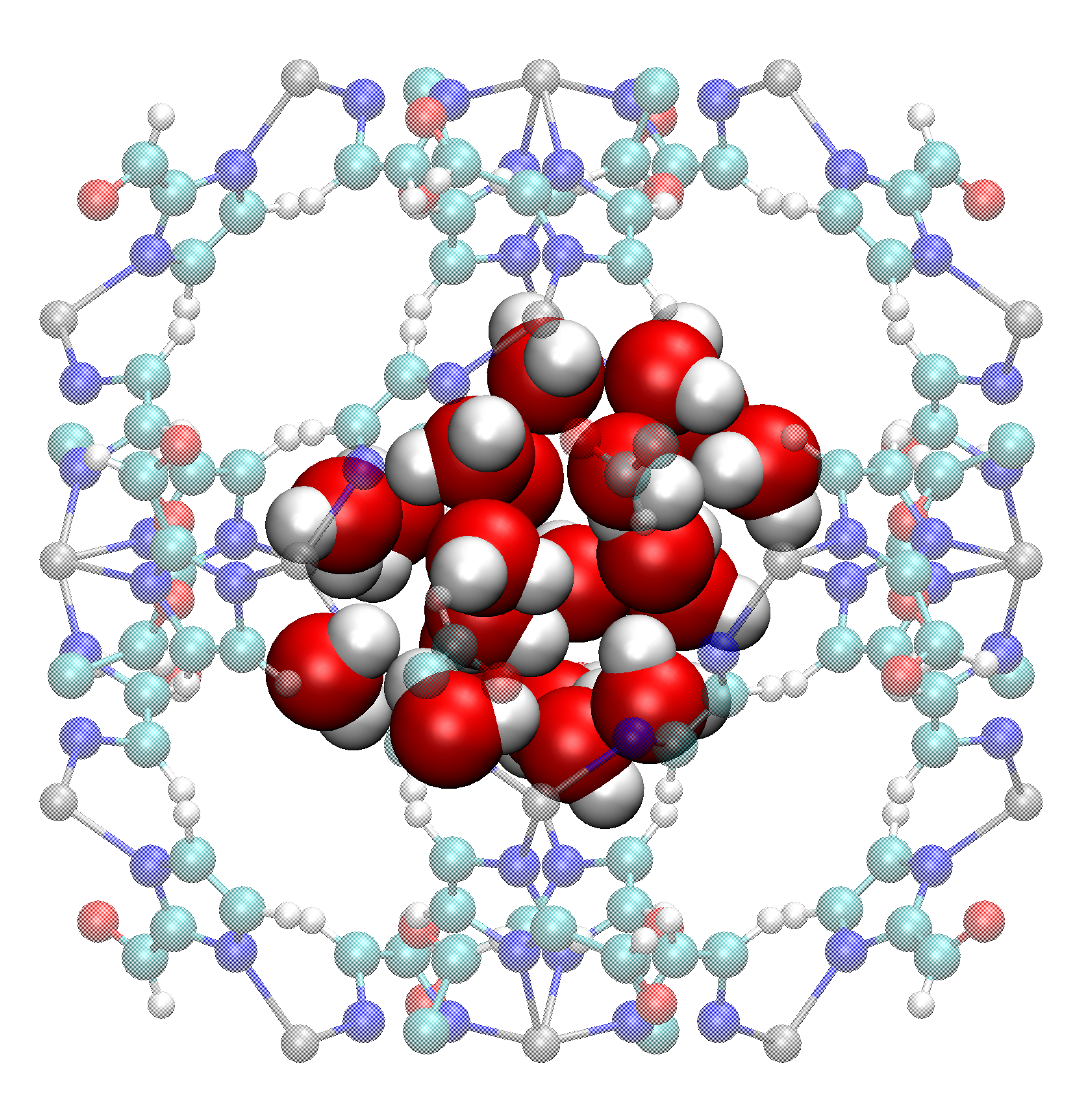}
\caption{Hybrid h-CMD method includes f-QCMD method for the light D$_2$O molecules inside the pores and f-CMD method for the ZIF-90 framework atoms. D$_2$O molecules in the adjacent pores have been excluded for visual clarity.}
\label{hybrid}
\end{figure}

In order to test the accuracy and applicability of our new h-CMD scheme, we use it to simulate IR spectra of interfacial D$_2$O molecules confined in the ZIF-90 framework. We specifically selected this system as experimental IR spectra were available for comparison. 
The natural division of this system in the h-CMD framework is to treat the D$_2$O molecules with f-QCMD and ZIF-90 with f-CMD, as depicted in Figure~\ref{hybrid}.

We would like to emphasize that this method allows for improved accuracy in terms of the removal of the curvature problem for the relevant molecules while being simple to implement for large, more complex, and heterogeneous condensed phases and interfaces. The h-CMD will suffer in accuracy for systems where spectral features of interest are part of the large molecule/framework that is being treated with f-CMD, as that feature will still exhibit the curvature problem. We still believe that the method has wide use for complex condensed phase and interfacial systems where the spectra of small molecule adsorbates are the focus.

\subsection{\label{sec2.1:ir}Infrared spectra calculations}

In this work, the infrared absorption spectrum is simulated using the cell dipole-moment derivative,
\begin{equation}
    n(\omega)\alpha(\omega) = \frac{\pi\beta}{3cV\epsilon_0}\tilde{I}(\omega),
\end{equation}
where $n(\omega)$ is the refractive index, $\alpha(\omega)$ is the absorption cross-section, and $\tilde{I}(\omega)$ 
is the Fourier transform of the total dipole-derivative autocorrelation function,
\begin{equation}
       \tilde{I}(\omega) = \frac{1}{2\pi}\int_{-\infty}^{\infty}dt\ \mathrm{e}^{-i\omega t} \langle
       \dot{\boldsymbol{\mu}}(0)\cdot\dot{\boldsymbol{\mu}}(t) \rangle f(t)
\end{equation}
where $\langle ... \rangle$ denote the microcanonical ensemble average, $\dot{\boldsymbol{\mu}}$ is the time
derivative of the total dipole moment of the system, and $f(t)$ is a window function that dampens the tail of the
autocorrelation function. Here we use the Hann window~\cite{press_etal:1992}
\begin{equation}
  f(t) = 
  \begin{cases}
    \cos^2\left( \frac{\pi t}{2\tau}\right)\quad &|t| \leq \tau \\
      0 & |t| > \tau
  \end{cases}  
\end{equation}
 where $\tau$ is a cutoff time chosen as appropriate.

\section{\label{sec:sim-det}Simulation details}

\subsection{\label{sec:init}System initialization}
All classical and quantum PI simulations are performed using a development version of our software package DL\_POLY  Quantum 2.0\cite{DLQ} using the experimentally obtained 1$\times$1$\times$1 crystal structure of ZIF-90,\cite{pnas_103_10186} utilizing periodic boundary conditions. DL\_POLY  Quantum 2.0 is a highly parallelized, modular, and user-friendly computational platform for classical and advanced PI simulations in condensed phases. The flexible anharmonic q-TIP4P/F\cite{Habershon:2009} quantum water potential was used for modeling water, which is shown to be very successful in reproducing different properties of water, including its phase diagram \cite{pccp_14_10140}, melting point\cite{pccp_13_19714, pccp_14_15199, jcp_133_144511}, and temperature of maximum density\cite{Habershon:2009, pccp_14_15199} as well as the magnitude of the isotope fractionations\cite{Markland7988} and vibrational infrared spectra\cite{Habershon:2009}. The equations of motion were propagated using the velocity--Verlet algorithm with a time step of 0.2 fs, with the temperature being kept constant at 300 K. The path integral Langevin equation (PILE) thermostat\cite{sepCeriotti2010} with a relaxation time of 0.05 ps and barostat with a relaxation time of 2.0 ps were employed as implemented in DL\_POLY Quantum 2.0. All bonded and non-bonded parameters of the ZIF-90 framework were obtained from Ref.\citenum{jpcc_125_12451} (see SI Table S1). Water--framework interactions were included using the non-bonded electrostatic and Lennard-Jones potentials. Lorentz--Berthelot mixing rules were used to drive cross-interaction terms with parameters for water taken from the q-TIP4P/F water model, which is the same as the original TIP4P/2005\cite{Abascal:2005}. 

For this study, we focused on the highest relative humidity (RH) of 60\%, which is also considered to be the saturation limit of this material at room temperature. The initial packing of D$_2$O molecules within the ZIF-90 framework was performed using PACKMOL\cite{Martnez:2009}, ensuring random distribution inside the cavities based on geometrical constraints.
To achieve a fully randomized initial configuration of the D$_2$O molecules within the ZIF-90 framework, simulated annealing was performed using three short consecutive canonical NVT simulations at varying temperatures. Initially, the system was simulated at 1000 K for 50 ps to facilitate rapid diffusion and avoid local clustering. Subsequently, the temperature was lowered to 500 K for 30 ps, and finally, the system was equilibrated at the target temperature of 300 K for 20 ps. This step-wise reduction in temperature was designed to ensure uniform distribution of the confined D$_2$O molecules. Atom-atom cutoff distances were set to 8.5~\AA~ to truncate the short-range interactions, whereas the long-range electrostatics were calculated using the Ewald summation technique.\cite{Leach:2001}
Following simulated annealing, the system was equilibrated in the isobaric-isothermal (NPT) ensemble at 300 K and 1 atm for 500 ps, allowing the volume to relax (See Supplementary Material Table S2 for the MD and PIMD equilibrated cell lengths compared to the experiment). The system was further equilibrated using 32-bead PIMD simulations in the NPT ensemble for 200 ps. PIMD equilibrations were then performed in the NVT ensemble for 50 ps using a PILE thermostat, followed by an additional 50 ps production simulation. 

\subsection{\label{sec3.2:sfg}Simulation Details for h-CMD}

For h-CMD simulations, we partition the system so that the adsorbed water molecules are treated with the f-QCMD method while the ZIF-90 framework atoms are treated with f-CMD as schematically depicted in Figure~\ref{hybrid}. Additionally, we further approximate our treatment of the framework atoms by setting all correction potentials for interactions involving just the framework atoms to zero. This reduces our intra-molecular corrections to be just for the water O-D bond and D-O-D angle and our inter-molecular corrections to be the water-water and water-ZIF interactions. We justify this treatment on several bases. Firstly, we are focused on the
O-D stretch region in our spectra calculation, so it is more important to correct for the curvature problem in the water molecules as opposed to the framework. Second, the framework is composed primarily of heavy atoms in which NQEs are
negligible at room temperature, meaning the correction potentials within the framework should be small enough to be ignored. Lastly, while there are hydrogen atoms within the framework that do exhibit NQEs, they do not strongly interact with the water molecules as the water molecules donate hydrogen bonds to the framework rather than accepting them. 

This treatment results in a total of 19 pair interactions, for which we calculate the correction potentials. To further test the importance of the different types of pair interactions, we also consider the case where we only correct the water-water pairs, leaving us with only 3 RDFs that are part of the inter-molecular corrections in the IBI process. To generate the distribution functions for the IBI process, we averaged over 50 independent PIMD simulations with 32 beads and an integration time step of 0.2 fs. The PILE thermostat with a time constant of 0.05 ps is used to achieve the NVT ensemble at 300 K. Each trajectory is first equilibrated for 50 ps followed by 150 ps of production with the distribution functions calculated every 100 time steps using histogram binning.

During the IBI process, the distribution functions are generated from classical-like simulations under the correction potentials. The regularization parameter, $\varepsilon$, is set to 10, and a total of 20 IBI iterations are performed to converge the correction potentials; see the Supplementary Material Figure S9 for the RMSE convergences with respect to the number of IBI iterations. For each iteration, the distribution functions were averaged over 50 independent NVT simulations. Each trajectory is evolved with a time step of 0.2 fs and thermostatted with a massive Nos\'{e}-Hoover chain (m-NHC) thermostat\cite{suzuki_32_400,YOSHIDA_150_262} of length 3 and a time constant of 0.05 ps. Trajectories were equilibrated for 50 ps and an additional 50 ps for data collection, where the distribution functions were calculated every 100 time steps with the RDFs calculated using force sampling. The correction potential updates for each iteration were calculated using a modified version of the IBI code provided by the authors of Ref.~\citenum{fqcmd_water_ice}.

For comparison, we also computed the vibrational spectra for the entire system using f-CMD. As with h-CMD, we only correct the water intramolecular potential and the water-water and water-framework intermolecular potentials. The simulation details are the same as the h-CMD discussed above but with the Cartesian centroid used in all distribution function calculations.

\subsection{Spectra simulation details}
The initial configurations for the real-time dynamics trajectories for the spectra calculations were taken from the final set of PIMD trajectories discussed in Section~\ref{sec:init}, with configurations saved every 1 ps, providing a total of 50 independent initial configurations for subsequent dynamics simulation using PA-CMD and T-RPMD methods.
Spectra calculations for all methods follow the same basic procedure. Configurations for dynamics trajectories are generated from NVT simulations at 300 K that involve 50 ps of equilibration and then 50 ps of sampling, with configurations taken every 1 ps. Sampling for h-CMD, and f-CMD are all done using the m-NHC thermostat with a chain length of 3 and a time constant of 0.05 ps. Sampling for other PI methods is done with PIMD simulations using 32 beads and the PILE thermostat with a time constant of 0.05 ps. Finally, sampling for the classical MD simulation is done similarly to the PI methods, except that a single-bead RP is used, effectively treating the system classically.

Real-time dynamics for all methods are performed with 50 independent NVE trajectories for 50 ps. PA-CMD simulations use 32 beads, and the internal RP modes are thermostatted at 300 K with the m-NHC thermostats of length 3. T-RPMD simulations are done with 32 beads, and the internal modes are thermostatted with the PILE thermostat. For simulations at the lowest temperature of 250 K, a total of 64 beads is used. IR spectra for D$_2$O are calculated using the total dipole derivative correlation function of the water molecules in the system. The first 5 ps of each trajectory is dropped, with the second 45 ps used. A Hann window with $\tau=1$ ps is applied to the correlation function.

\section{\label{sec:results}Results and discussion}
As mentioned, IR spectroscopy is an excellent tool for probing the complex HB network of water in different condensed phase and interface environments. Experimentally, the deep inelastic neutron scattering (DINS) technique can provide direct information on proton momentum distribution and has been applied to both liquid bulk water\cite{prl_100_177801} and water confined in hydrophobic Nafion\cite{prl_111_036803} with average pore diameters of $\sim$20~\AA, silica nanopores\cite{jcp_127_154501} with average pore diameters of $\sim$24~\AA~and $\sim$82~\AA~as well as to carbon nanotubes with $\sim$16~\AA~ diameters\cite{prb_85_045403}. The data from DINS on water confined in Nafion showed that the nanoconfined water is in a distinctly different quantum state than that of bulk water.\cite{prl_111_036803} From the theory standpoint, despite recent theoretical studies attempting to provide molecular-level insights into water structures at different interfaces, the impact of NQEs on the vibrational spectral signatures of the interfacial confined water still remains an open question. Here, we use the complex system of interfacial D$_2$O molecules confined in the ZIF-90 framework to test the accuracy of our new h-CMD method for infrared spectral simulations of large complex systems compared to a myriad of classical and quantum PI simulation methods as implemented in our in-house developed software DL\_POLY Quantum 2.0. We then discuss the ability of each approach to predict the infrared spectra compared to the available experiment at different temperatures of 250-600 K.

\subsection{\label{sec4.3:ir}The Test case of confined D$_2$O in the extended ZIF-90 MOF.}

\begin{figure}[t]
\centering
\includegraphics[width=0.99\linewidth]{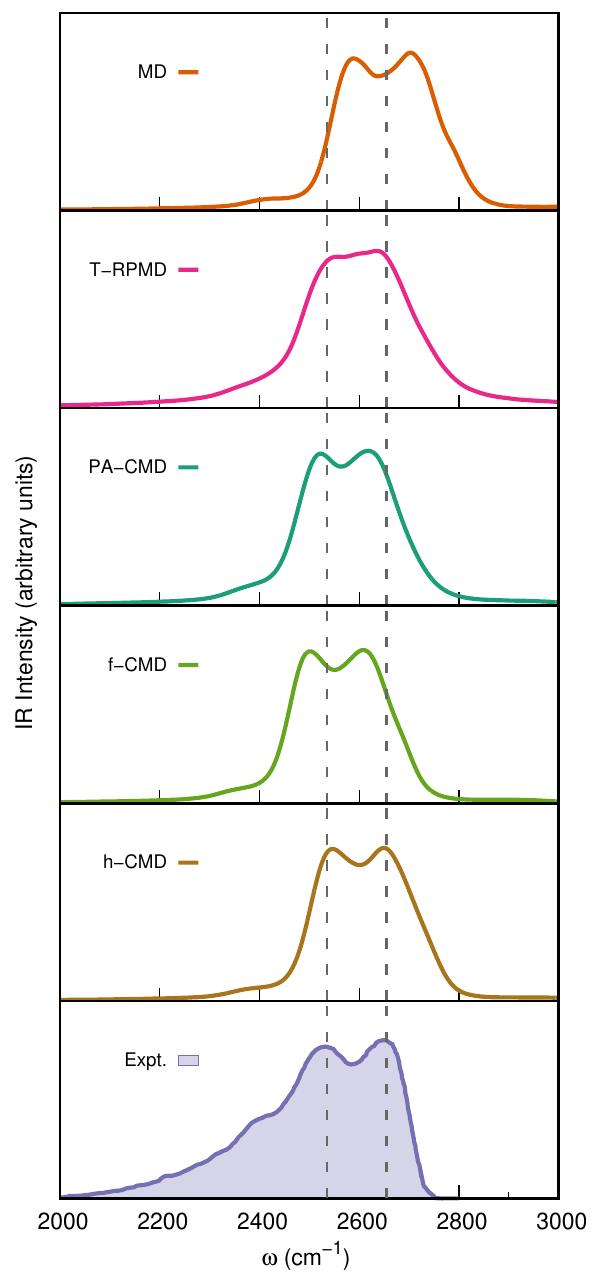}
\caption{Classical MD, T-RPMD, PA-CMD, and f-CMD simulated IR spectra of confined D$_2$O in ZIF-90 in the O--D stretch region at 60\% RHs and at 300 K compared to that of h-CMD and experiment.\cite{jpcc_125_12451} The corresponding deconvoluted spectra are provided in the Supplementary Material Figure S3.}
\label{spectra-zif90}
\end{figure}

As mentioned, here we aim to test the accuracy of our h-CMD scheme for the simulation of the vibrational spectra of D$_2$O confined in ZIF-90 compared to other approximate PI methods and the experiment as reference (Figure \ref{spectra-zif90}).\cite{jpcc_125_12451} 
We note that an accurate description of the interatomic potentials is important for capturing the complex water-water vs. water-framework interactions at the interface. To reproduce the correct spectral positions and line shapes as obtained from the experiment, the non-bonded Lennard-Jones (LJ) potential parameters between the carbonyl oxygen of the framework and the oxygen atoms of the D$_2$O adsorbate molecules were fine-tuned. After detailed analyses, the modified force field parameters were used to calculate the IR spectra of confined D$_2$O in ZIF-90 using different PI flavors, including the newly introduced h-CMD method (See the Supplementary Material Section S1 and Table S1 and Figure S2 for more details). In Ref. \citenum{jpcc_125_12451}, for all considered \%RHs, a doublet feature was observed, which was attributed to the bulk and interfacial water, corresponding to the low- and high-frequency peaks, respectively, in Figure \ref{spectra-zif90}. The interested reader is referred to the Supplementary Material Figure S3 for detailed analyses of these peaks. While classical MD simulations provide the characteristic doublet feature of the confined water, they grossly underestimate the strength of the HB network due to neglecting NQEs. As a result, both peaks are blue-shifted in the considered O--D stretch region.
We note that the missing intensity in the 2400 cm$^{-1}$ region from all the simulated spectra compared to the experiment is due to the Fermi resonances, which neither MD nor any of the PI methods are able to capture quantitatively.\cite{fermi-resonance}  

Considering the lower frequency HB peak from the experiment, which appears at 2535 cm$^{-1}$, our classical MD simulated spectrum shows a blue shift of 56 cm$^{-1}$. T-RPMD reduces this blue shift, positioning the first peak at the experimental value, but fails to accurately resolve the doublet feature due to the inherent artificial broadening of spectra caused by excessive friction in the dynamics. On the other hand, the PA-CMD calculated spectrum exhibits a red-shift of 12 cm$^{-1}$ for the first peak compared to the experiment stemming from its well-known curvature problem. In contrast, our newly introduced h-CMD method successfully reproduces the doublet feature demonstrating this method's ability to overcome the curvature problem with its O--D peak positioning in line with T-RPMD. Meanwhile, f-CMD behaves rather similarly to PA-CMD with slightly more red-shifted O--D peaks, a behavior which was also observed in bulk water spectra for the force-matching f-CMD method.~\cite{mayYuan2018}  

The second higher-frequency experimental peak, which corresponds to the interfacial water molecules interacting with the ZIF-90 framework, appears at 2655 cm$^{-1}$. Classical MD once again demonstrates a notable blue shift of 49 cm$^{-1}$, reflective of its limitations in capturing the subtleties of HBs in confined water. As mentioned, with smeared peak structures, the fine details of the spectrum is lost in T-RPMD simulated spectra while PA-CMD red-shifts the second peak by 36 cm$^{-1}$. On the other hand, h-CMD accurately captures both peak positions, with minimal shifts from the experimental data, and successfully resolves the doublet feature. The close agreement in both the positions and the relative intensities of the peaks between h-CMD and the experiment highlights its accuracy for calculating vibrational spectra while incorporating NQEs.

\subsection{\label{sec4.3:ir}Approximate h-CMD.}
\begin{figure}[!t]
\centering
\includegraphics[width=0.99\linewidth]{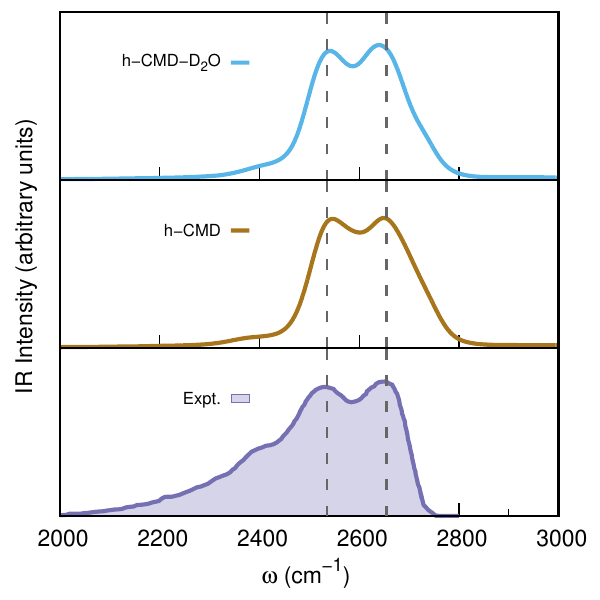}
\caption{h-CMD simulated IR spectra of confined D$_2$O in ZIF-90, with h-CMD-D$_2$O with only 3 inter-molecular water-water pairs corrections compared to that of full h-CMD and experiment.\cite{jpcc_125_12451}}
\label{spectra-zif90-hcmd3}
\end{figure}

As mentioned in Section \ref{sec3.2:sfg}, in h-CMD, intra-molecular corrections are considered only for the water O-D and D-O-D bonds and angles, with inter-molecular corrections included for all the water-water and water-framework interactions.
Here, we consider the case of h-CMD with further simplification to the inter-molecular correction potentials. In this approximate h-CMD method, which we have labeled as h-CMD-D$_2$O, we only include correction potentials for the 3 water-water pairs. The resulting IR spectra for h-CMD-D$_2$O is presented in Figure~\ref{spectra-zif90-hcmd3}, where it is compared to the case with all water-water and water-framework interactions corrected as well as the experimental spectrum as reference.\cite{jpcc_125_12451}
As expected, the full h-CMD method, which considers all inter-molecular interactions, results in an IR spectrum that closely resembles that of the experiment. Interestingly, the h-CMD-D$_2$O method does provide a spectrum that is very close to the full h-CMD method, with only minor red-shifting of the higher frequency peak. This is as expected since this peak corresponds to the water molecules donating HBs to the carbonyl oxygen of the ZIF-90 framework, and this interaction has not been corrected. However, this result does indicate that for a given system, the number of correction potentials needed can be optimized to reduce the computational cost without sacrificing too much accuracy.

\subsection{\label{sec4.3:temp-depen}Temperature dependence of IR spectra of confined D$_2$O in ZIF-90.}

\begin{figure}[!h]
\centering
\includegraphics[width=0.99\linewidth]{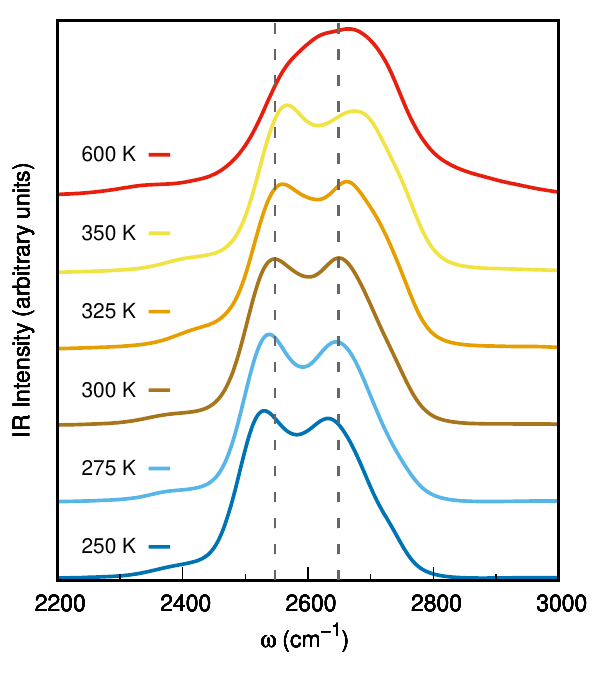}
\caption{Temperature--dependent h-CMD simulated IR spectra of confined D$_2$O in ZIF-90 in the O--D stretch region. Two vertical dashed lines represent peak positions at 300 K.}
\label{ir-temp}
\end{figure}

Here, we explore the temperature-dependent IR spectral line shapes and positions over a range of different temperatures from 250-600 K for confined D$_2$O in ZIF-90 as calculated from the h-CMD method. As can be seen from Figure~\ref{ir-temp}, the IR spectra are red-shifted upon lowering the temperature compared to the 300 K spectra and blue-shifted as the temperature increases. 
As the temperature decreases, the slight red-shift of the low-frequency HB peak could be primarily due to the slowdown of the molecular motions of the confined D$_2$O molecules, allowing for stronger and more stable HBs. At the same time, some of these small red shifting could be artificial and due to the curvature problem starting to show itself at lower temperatures. The amount of the red-shift is, however, less noticeable in the case of the higher frequency interfacial peak. On the other hand, as the temperature increases, the calculated IR peaks blue-shift as the provided extra thermal energy increases molecular motions, leading to the weakening of the HBs and overall broadening and loss of the doublet feature at the highest studied temperature of 600 K.
This behavior is also observed in QCMD simulations of bulk water.~\cite{qcmd}
Also, the lower frequency peak associated with the bulk-like water, slightly increases in relative intensity as the temperature decreases. This could signal that lowering the temperature leads to confined D$_2$O molecules forming more organized structures with themselves similar to bulk water, increasing the population of molecules in bulk-like configurations within the ZIF-90 pores. This increase in HB stabilization promotes bulk-like characteristics in the spectra, enhancing the relative intensity of the first peak. This trend reflects the shift from more surface or interfacial water at higher temperatures to more bulk-like configurations at lower temperatures within the confined space.
Thus, these spectral changes with temperature highlight the delicate balance between hydrogen bonding and thermal energy in determining the structural and vibrational behavior of confined water in nanoscopic environments.

\begin{figure}[!h]
\centering
\includegraphics[width=0.99\linewidth]{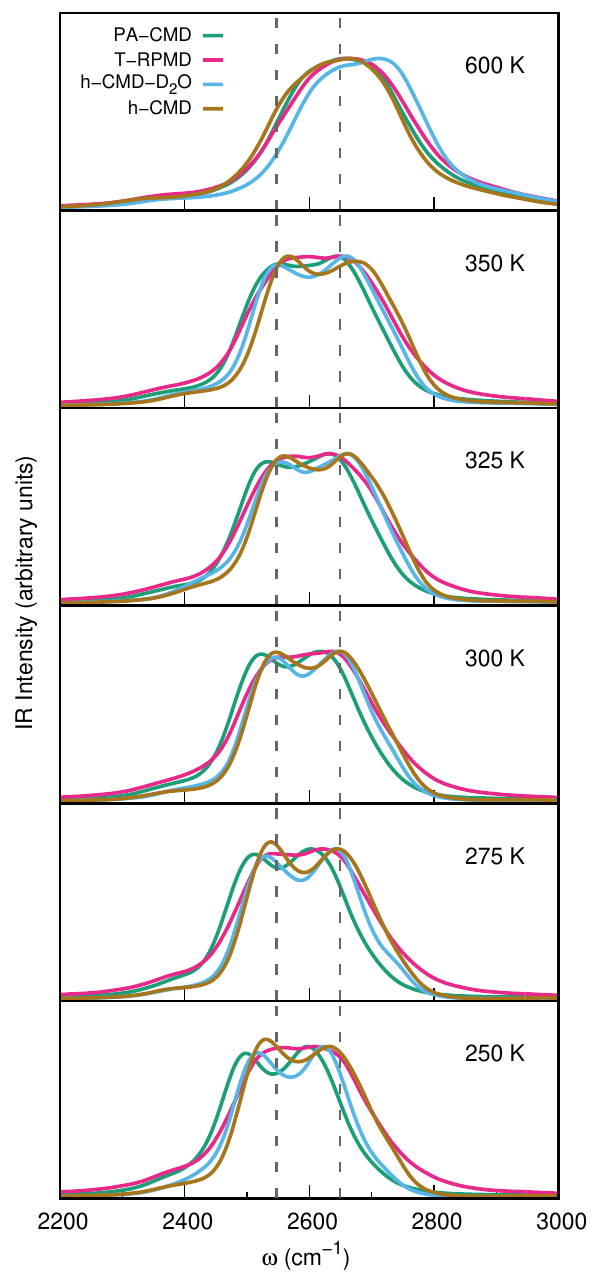}
\caption{Temperature--dependent IR spectra of confined D$_2$O in ZIF-90 in the O--D stretch region using different methods. Two vertical dashed lines represent h-CMD calculated peak positions at 300 K.}
\label{ir-temp-all}
\end{figure}

Figure~\ref{ir-temp-all} compares the temperature dependence of the IR spectra for PA-CMD and T-RPMD methods compared to that of h-CMD and h-CMD-D$_2$O. For the highest studied temperature of 600 K, except for h-CMD-D$_2$O, the simulated spectra calculated using all methods match very closely. All methods show a shifting of the spectra to lower frequencies with lowering temperatures. This shifting is the most dramatic with PA-CMD, highlighting how the curvature problem worsens with lowering the temperature, even for the heavier isotope of water. The broadening of the features of T-RPMD spectra is also demonstrated, with no doublet feature appearing at every temperature. The last comparison of notes is between h-CMD and h-CMD-D$_2$O. While the relative peak intensities do vary between the two treatments, the peak positions are very close to each other, except in the 600 K case, further showcasing the tunability of our scheme in reducing the number of correction potentials.

\section{\label{sec5:tune}Conclusion}
In this work, we have introduced a new hybrid f-CMD/f-QCMD scheme, where the accuracy of the computationally expensive f-QCMD method is reserved for molecules that are believed to suffer from the curvature problem more severely while the rest of the system is treated with the computationally less expensive f-CMD method where determining the quasi-centroid coordinates is difficult or is entirely unfeasible. By simulating IR spectra of D$_2$O molecules confined in the extended ZIF-90 framework as a test case, the efficiency and accuracy of this scheme are showcased for large-scale PI simulations of complex condensed phase and interfacial systems. Our detailed analyses show that the newly introduced h-CMD method is capable of capturing the twin lower and higher frequency bulk-like and interfacial peaks for the nanoconfined water in ZIF-90 compared to the experiment. This is in sharp contrast to the traditionally employed T-RPMD method, where the artificial broadening of the IR spectra leads to the loss of the doublet feature, especially as the temperature is lowered. Future works will focus on combining our h-CMD scheme with accurate neural network potentials for predictive simulations of the vibrational spectra of the complex heterogeneous condensed phase and interfacial water in different media.

\section*{Supplementay Material}
Convergence analysis of intra- and intermolecular RDFs, deconvoluted IR spectra for various computational methods, and details of the potential parameters used.

\begin{acknowledgments}
This research was partially supported by the National Science Foundation through award no. CBET-2302618. Simulations used resources from Bridges2 at Pittsburgh Supercomputing Center through allocation PHY230099, PHY240170, CHE240103 from the Extreme Science and Engineering Discovery Environment (XSEDE),\cite{xsede} which was supported by National Science Foundation grant number 1548562. The use of computing resources and support provided by the HPC center at UMKC is also gratefully acknowledged.
\end{acknowledgments}

\section*{\label{sec_contrib} AUTHOR DECLARATIONS}
$^{\dagger}$D.K.L. and N.L. contributed equally to this work.

\subsection*{Conflict of Interest}

\noindent The authors have no conflicts to disclose.

\subsection*{Author Contributions}
\textbf{Dil K. Limbu}: Software (equal); h-CMD Method (supporting); Writing - original draft (supporting); Writing - review \& editing (equal). 
\textbf{Nathan London}: Software (equal); h-CMD Method (equal); Writing - original draft (equal); Writing - review \& editing (equal). 
\textbf{Md Omar Faruque}: Software (supporting); h-CMD Method (supporting); Writing - original draft (supporting); Writing - review \& editing (equal).
\textbf{Mohammad R. Momeni}: Conceptualization (lead); Funding acquisition (lead); Supervision (lead); h-CMD Method (equal); Software (supporting); Writing - original draft (equal); Writing - review \& editing (equal). 

\section*{\label{sec_contrib} DATA AVAILABILITY}
The data that support the findings of this study is available from the corresponding authors upon reasonable request. The DL\_POLY Quantum software package
is available for download at https://github.com/dlpolyquantum/dlpoly\_quantum

\section*{REFERENCES}
\bibliography{articles,software}


\end{document}


$^{\dagger}$D.K.L. and N.L. contributed equally to this work.

\newpage
\tableofcontents
\newpage

\clearpage
\noindent \textbf{Section S1. Fine-tuning force field parameters and its influence on IR line shape}
\addcontentsline{toc}{section}{Section S1. Fine-tuning force field parameters}

A key challenge in accurately simulating the IR spectra of confined water is capturing the correct interactions between the water molecules and the framework in which they are confined. In this work, we modify the non-bonded Lennard-Jones (LJ) parameters between the carbonyl oxygen (O) of the ZIF-90 framework and the oxygen atoms of D$_2$O (O$_W$) by reducing $\sigma_{O-O_W}$ by 10\% and increasing $\epsilon_{O-O_W}$ by 50\%. This adjustment reduces the O--O$_W$ repulsion, allowing D$_2$O molecules to come closer to the framework and form stronger HBs with the carbonyl groups~\cite{jpcc_125_12451}, which is crucial in reproducing the correct IR line shapes observed in experimental IR spectra.
\begin{figure}[!h] 
\centering 
\includegraphics[width=0.9\textwidth]{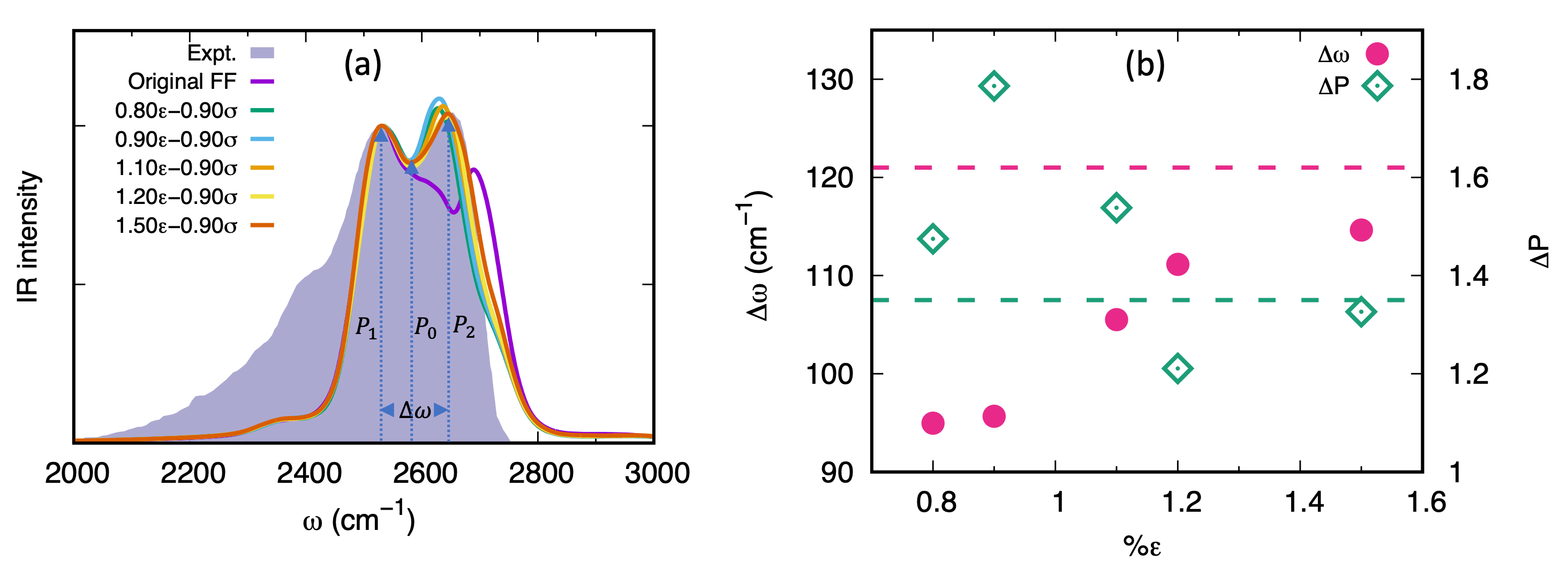} 
\caption{(a) Influence of the modified force field parameters for the O of the carbonyl of the framework to the oxygen of water (O--O$_W$) interactions on the classical MD calculated IR line shape in the O--D stretch region. The panel shows variations in the IR spectra resulting from different values of $\epsilon$ for a 10\% decrease in $\sigma$ for the O--O$_W$ interaction in the hydrated ZIF-90 system at 60\% RH. $\Delta \omega$ in panel (b) represents the separation between the two peaks, while $\Delta P$ corresponds to the ratio of the intensity of the second peak (P$_2$) to the first peak (P$_1$), measured from the minimum intensity in the middle ($P_0$). Dashed lines in panel (b) indicate experimental values. The modified force field with $\epsilon$ increased by 50\% (1.5$\epsilon$) and $\sigma$ decreased by 10\% (0.9$\sigma$), captures the water-hydroxy interaction in the O--D stretch region. All classical MD calculated IR spectra are red-shifted by 57 cm$^{-1}$ for easier comparisons to the experiment.}
\label{fig-fit}
\addcontentsline{toc}{subsection}{Figure \ref{fig-fit}. Influence of modified FF parameters on the calculated IR line shapes.}
\end{figure}

\clearpage
\begin{table}
\caption{Force field parameters for ZIF-90.~\cite{jpcc_125_12451}}
\addcontentsline{toc}{subsection}{Table \ref{table_ff}. Force field parameters for ZIF-90.}
    \begin{subfigure}{.5\textwidth}
    \centering
    \includegraphics[width=.6\linewidth]{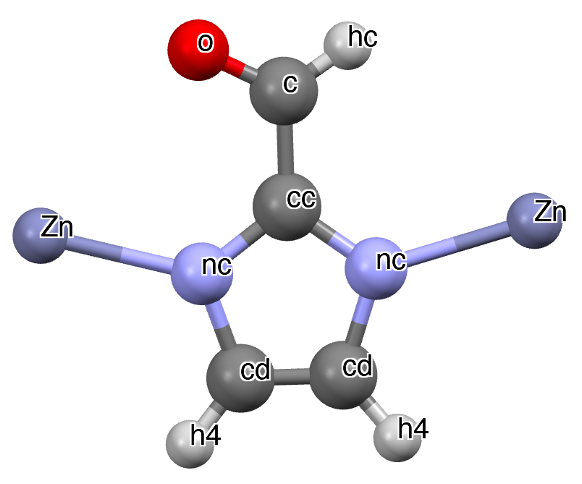} 
    \end{subfigure}
    
    \begin{subtable}{0.49\textwidth}
    \centering
    \caption{\textbf{Non-bonded} potential terms.}
    \scalebox{0.85}{
    \begin{tabular}{c c c c c}
       \hline
       \specialcell{Atom\\name} & \specialcell{Atom\\type} & \specialcell{q\\(e)} & \specialcell{$\epsilon$\\(kcal/mol)} & \specialcell{$\sigma/2$\\({\AA})} \\
       \hline
        Zn     & Zn   & 0.6428   & 0.01251  & 0.9800  \\
        nc     & nc   & -0.2993  & 0.17010  & 1.6250 \\
        o      & o    & -0.3475  & 0.21000  & 1.6612 \\
        c      & c    & 0.1864   & 0.08605  & 1.7000 \\
        cc     & cc   & 0.1937   & 0.08605  & 1.7000 \\
        cd     & cd   & -0.0062  & 0.08605  & 1.7000  \\
        h4     & h4   & 0.1059   & 0.01501  & 1.2550 \\ 
        hc     & hc   & 0.0452   & 0.01571  & 1.2355 \\ 
        \hline
        \end{tabular}}
    \end{subtable}%
    \begin{subtable}{0.49\textwidth}
    \centering
        \caption{\textbf{Bond potential}: $U(r) = \frac{1}{2}K_r(r-r_0)^2$}
        \scalebox{0.85}{
        \begin{tabular}{c c c}
        \hline
        \textbf{Bond} & \specialcell{$K_r$ \\ (kcal.mol$^{-1}${\AA}$^2$)} & \specialcell{r$_0$ \\ ({\AA})} \\
        \hline
        Zn--nc    & 149.00   & 2.024 \\
        nc--cc    & 882.20   & 1.369  \\
        nc--cd    & 1050.80   & 1.317  \\
        cc--c     & 742.00   & 1.468   \\
        c--hc    & 621.40   & 1.112  \\
        c--o     & 1275.40   & 1.218  \\
        cd--cd    & 839.60   & 1.428  \\
       cd--h4    & 650.00   & 1.082  \\
        \hline
        \end{tabular}}
    \end{subtable}
    \medskip

    \begin{subtable}{0.49\textwidth}
    \centering
    \caption{\textbf{Angle potential}: $U(\theta) = \frac{1}{2}K_\theta(\theta-\theta_0)^2$}
    \scalebox{0.85}{
    \begin{tabular}{c c c}
        \hline
        \textbf{Angle} & \specialcell{$K_\theta$ \\ (kcal.mol$^{-1}$.rad$^2$)} & \specialcell{$\theta_0$ \\ ( \degree~)}\\
        \hline
        Zn--nc--cc     & 28.8600    & 126.8500 \\
        Zn--nc--cd     & 22.7200    & 126.9500 \\
        nc--Zn--nc     & 2.1800    & 109.4200 \\
        nc--cc--nc     & 139.6000    & 125.7001 \\
        nc--cc--c      & 132.4000    & 123.3201 \\
        nc--cd--cd     & 143.2000    & 112.5600 \\
        nc--cd--h4     & 104.2000    & 118.4701 \\
        cc--nc--cd     & 143.6000    & 105.4900 \\
        cc--c--hc     & 94.2000    & 114.8300 \\
        cc--c--o      & 139.2000    & 123.9301 \\
        hc--c--o      & 108.4000    & 120.7001 \\
        cd--cd--h4     & 91.8000    & 127.9601 \\
        \hline
    \end{tabular}} 
    \end{subtable}%
    \begin{subtable}{0.49\textwidth}
    \centering
        \caption{\textbf{Torsion:} $U(\phi) = K_\phi[1+\cos(n\phi-\phi_0)]$}
        \scalebox{0.85}{
        \begin{tabular}{c c c c}
        \hline
        \textbf{Dihedral} & \specialcell{$K_\phi$ \\ (kcal.mol$^{-1}$)} & \specialcell{$\phi_0$ \\ ( \degree~)} & \specialcell{n \\ ~}\\
        \hline
        Zn--nc--cd--h4     & 1.056   & 180   & 2 \\
        Zn--nc--cd--cd     & 1.416   & 180   & 2 \\
        Zn--nc--cc--c      & 0.227   & 180   & 2 \\
        Zn--nc--cc--nc     & 0.614   & 180   & 2 \\
        nc--Zn--nc--cc     & 0.026   & 0   & 3 \\
        nc--Zn--nc--cd     & 0.021   & 180   & 3 \\
        nc--cc--nc--cd     & 4.750   & 180   & 2 \\
        nc--cc--c--hc     & 2.875   & 180   & 2 \\
        nc--cc--c--o      & 2.875   & 180   & 2 \\
        nc--cd--cd--nc     & 4.000   & 180   & 2 \\
        nc--cd--cd--h4     & 4.000   & 180   & 2 \\
        cc--nc--cd--cd     & 4.750   & 180   & 2 \\
        cc--nc--cd--h4     & 4.750   & 180   & 2 \\
        c--cc--nc--cd     & 4.750     & 180   & 2 \\
        h4--cd--cd--h4     & 4.000   & 180   & 2 \\
        \hline
        \end{tabular}}
    \end{subtable}
    \medskip

    \begin{subtable}{0.49\textwidth}
    \centering
        \caption{\textbf{Improper:} $U(\phi) = K_\phi[1+\cos(n\phi-\phi_0)]$}
        \scalebox{0.85}{
        \begin{tabular}{c c c c}
        \hline
        \textbf{Improper} & \specialcell{$K_\phi$ \\ (kcal.mol$^{-1}$)} & \specialcell{$\phi_0$ \\ ( \degree~)} & \specialcell{n \\ ~}\\
        \hline
        nc--Zn--cc--cd     & 0.056   & 180   & 2 \\
        cc--nc--c--nc      & 1.100   & 180   & 2 \\
        c--hc--cc--o       & 10.500  & 180   & 2 \\
        cd--h4--cd--nc     & 1.100   & 180   & 2 \\
        \hline
        \end{tabular}}
    \end{subtable}%
    \begin{subtable}{0.49\textwidth}
    \centering
    \caption{Fine-tuned \textbf{non-bonded} parameters.}
    \scalebox{0.85}{
    \begin{tabular}{c c c}
       \hline
        \specialcell{Atom\\type pair} & \specialcell{$\epsilon$\\(kcal/mol)} & \specialcell{$\sigma/2$\\({\AA})} \\
       \hline
        o-O$_w$   & 0.29582  & 1.4583  \\
        \hline
        \end{tabular}}
    \end{subtable}%
    \label{table_ff}
\end{table}

The initial simulations using the original FF display significant discrepancies from experimental IR spectra, particularly in the O-D stretch region. With the fine-tuned FF, however, we observe a closer alignment with experimental results, particularly in the O-D stretching region, as seen in Figure~\ref{fig-fit}. The fine-tuned FF significantly improves the accuracy by enhancing the strength and directionality of the framework--D$_2$O interactions. The spectrum reveals two distinct peaks: a lower-frequency peak ($\sim$2530 cm$^{-1}$) corresponding to bulk-like hydrogen bonds and a higher-frequency peak ($\sim$2650 cm$^{-1}$) corresponding to interfacial hydrogen bonds. These features are accurately captured after the force field modification.

In the fine-tuning process, we systematically explore how varying the LJ parameters $\epsilon$ and $\sigma$ affects the IR line shape of the confined D$_2$O molecules within the ZIF-90 framework. The resulting IR spectra were analyzed based on two critical metrics: The peak separation ($\Delta \omega$), which reflects the distance between the two distinct peaks in the O--D stretch region. This separation provides insights into the differentiation between bulk-like and interfacial hydrogen bonds within the confined D$_2$O. The second metric, the intensity ratio, $\Delta P$, measures the intensity of the second peak (P$_2$) relative to the first peak (P$_1$), which was calculated from the minimum intensity between them. This ratio is sensitive to the strength of the HB and the distribution of HBs across the bulk and interfacial environments. Through this systematic analysis, we observed that the fine-tuned force field parameters effectively capture the experimental line shape for the considered 60\% RH.

\clearpage
\begin{figure}[!h]
\centering
\includegraphics[width=0.45\linewidth]{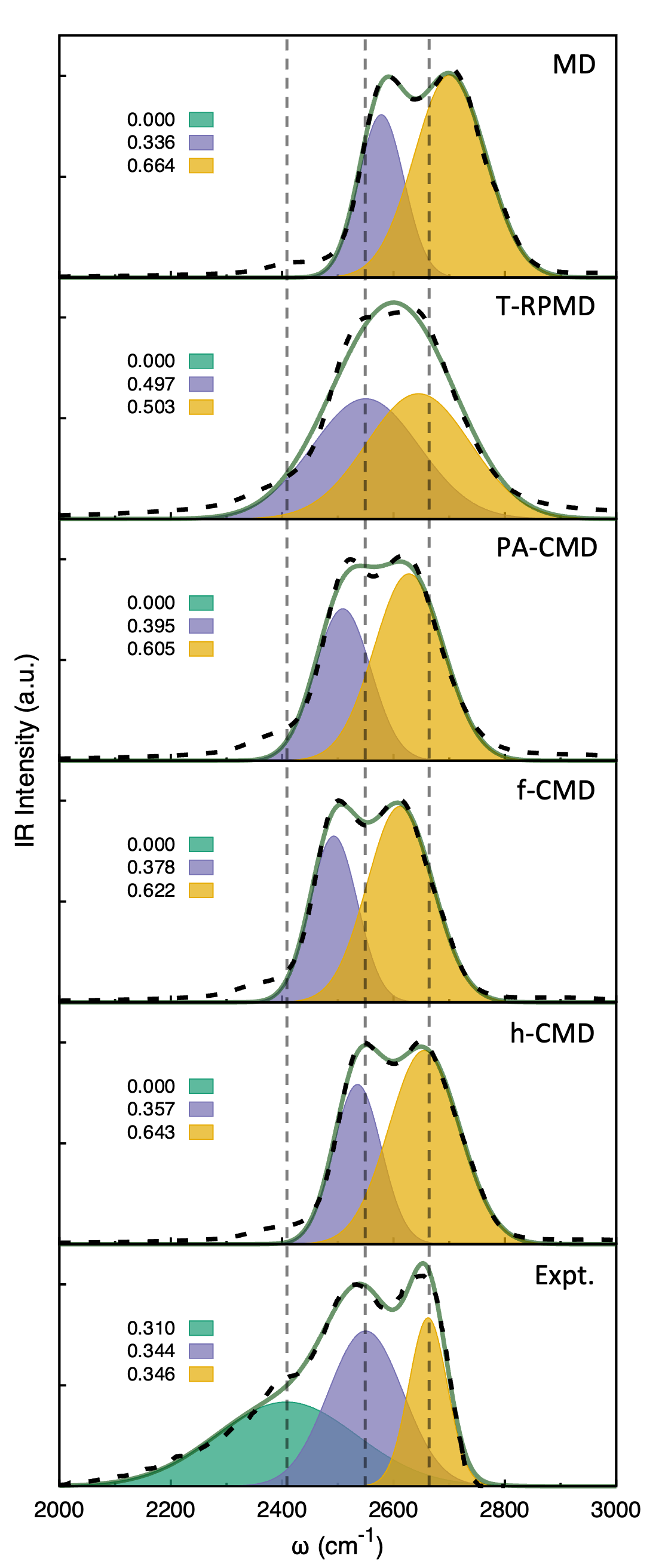}

\setlength{\tabcolsep}{1.2mm}
\scalebox{0.8}{
\begin{tabular}{c c c c c c c c c c c c c c c c c c c }
\midrule[1.4]
& \multicolumn{15}{c}{\textbf{ZIF-90}}  \\
\cline{2-19}
& \multicolumn{3}{c}{\textbf{MD}} & \multicolumn{3}{c}{\textbf{TRPMD}} & \multicolumn{3}{c}{\textbf{PA-CMD}} & \multicolumn{3}{c}{\textbf{f-CMD}} & \multicolumn{3}{c}{\textbf{h-CMD}} & \multicolumn{3}{c}{\textbf{Expt.}}  \\
\cline{2-19}
 Gaussian & $a_i$ & $\omega_i$ & $\sigma_{\omega_i}$ & $a_i$ & $\omega_i$ & $\sigma_{\omega_i}$ & $a_i$ & $\omega_i$ & $\sigma_{\omega_i}$ & $a_i$ & $\omega_i$ & $\sigma_{\omega_i}$ & $a_i$ & $\omega_i$ & $\sigma_{\omega_i}$ & $a_i$ & $\omega_i$ & $\sigma_{\omega_i}$ \\
\hline
$g_1$ & -- & -- & -- & -- & -- & -- & -- & --  & -- & -- & --  & -- & -- & --  & -- & 0.417 & 2408  & 126  \\
$g_2$ & 0.807  & 2578 & 41  & 0.597 & 2550 & 95 & 0.755 & 2509 & 50 & 0.824 & 2493 & 42 & 0.792 & 2535 & 43 & 0.769 & 2550 & 67 \\
$g_3$ & 1.005  & 2701 & 64  & 0.622 & 2645 & 93 & 0.928 & 2628 & 62 & 0.971 & 2612 & 59 & 0.961 & 2655 & 63 & 0.834 & 2662 & 34 \\
\midrule[1.4]
\end{tabular}}
\caption{Deconvolution of the main components of the O--D stretch of D$_2$O in ZIF-90 for different PI methods compared to that of the experiment\cite{jpcc_125_12451}.}
\addcontentsline{toc}{subsection}{Figure \ref{gaussian-fit}. Deconvolution of the O--D stretch for different considered methods.}
\label{gaussian-fit}
\end{figure}

\clearpage
\begin{table}
\caption{Average length of the cubic box of 1$\times$1$\times$1 unit cell for dry and hydrated ZIF--90 system as obtained from NPT MD and PIMD simulations compared to that of the experiment (expt.).}
\addcontentsline{toc}{subsection}{Table \ref{box_sizes}. Average simulated box length of ZIF--90 system}
\centering
\setlength{\tabcolsep}{2mm}
\renewcommand{\arraystretch}{1.5}
\scalebox{1.0}{
\begin{tabular}{c c c}
\midrule[1.4]
\multirow{2}{*}{Water Loading} & \multicolumn{2}{c}{Box Size ({\AA})}  \\
\cline{2-3}
 & MD & PIMD \\
\midrule[1.4]
dry ZIF--90 {\footnotesize(expt.)}  & \multicolumn{2}{c}{17.2715(4)} \\
dry ZIF--90  & 17.484 \pm 0.052  & 17.478\pm 0.069 \\
60\% RH  & 17.320 \pm 0.056 & 17.352\pm 0.089 \\
\midrule[1.4]
\end{tabular}}     
\label{box_sizes}
\end{table}

\clearpage
\begin{figure}[!ht] 
\centering 
\includegraphics[width=0.99\textwidth]{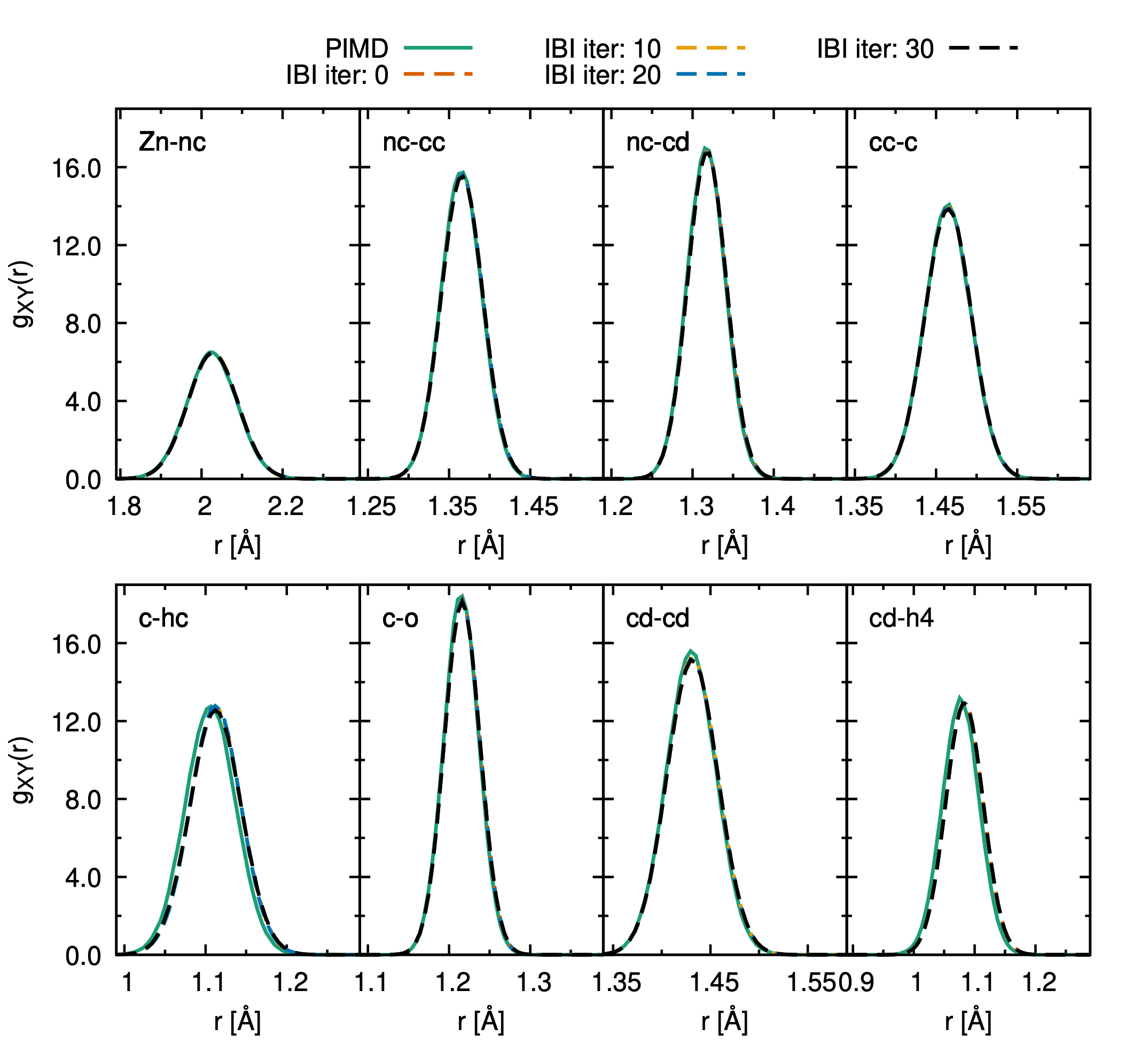} 
\caption{Comparison of intra-molecular bond lengths of ZIF-90 system between the IBI-corrected simulation and exact PIMD results across IBI iterations. The plot shows the convergence of bond lengths towards the exact PIMD values as the number of IBI iterations increases.}
\label{fig-zif-bond}
\addcontentsline{toc}{subsection}{Figure \ref{fig-zif-bond}. Intra-molecular bond lengths of ZIF-90.}
\end{figure}

\clearpage
\begin{figure}[!ht] 
\centering 
\includegraphics[width=0.99\textwidth]{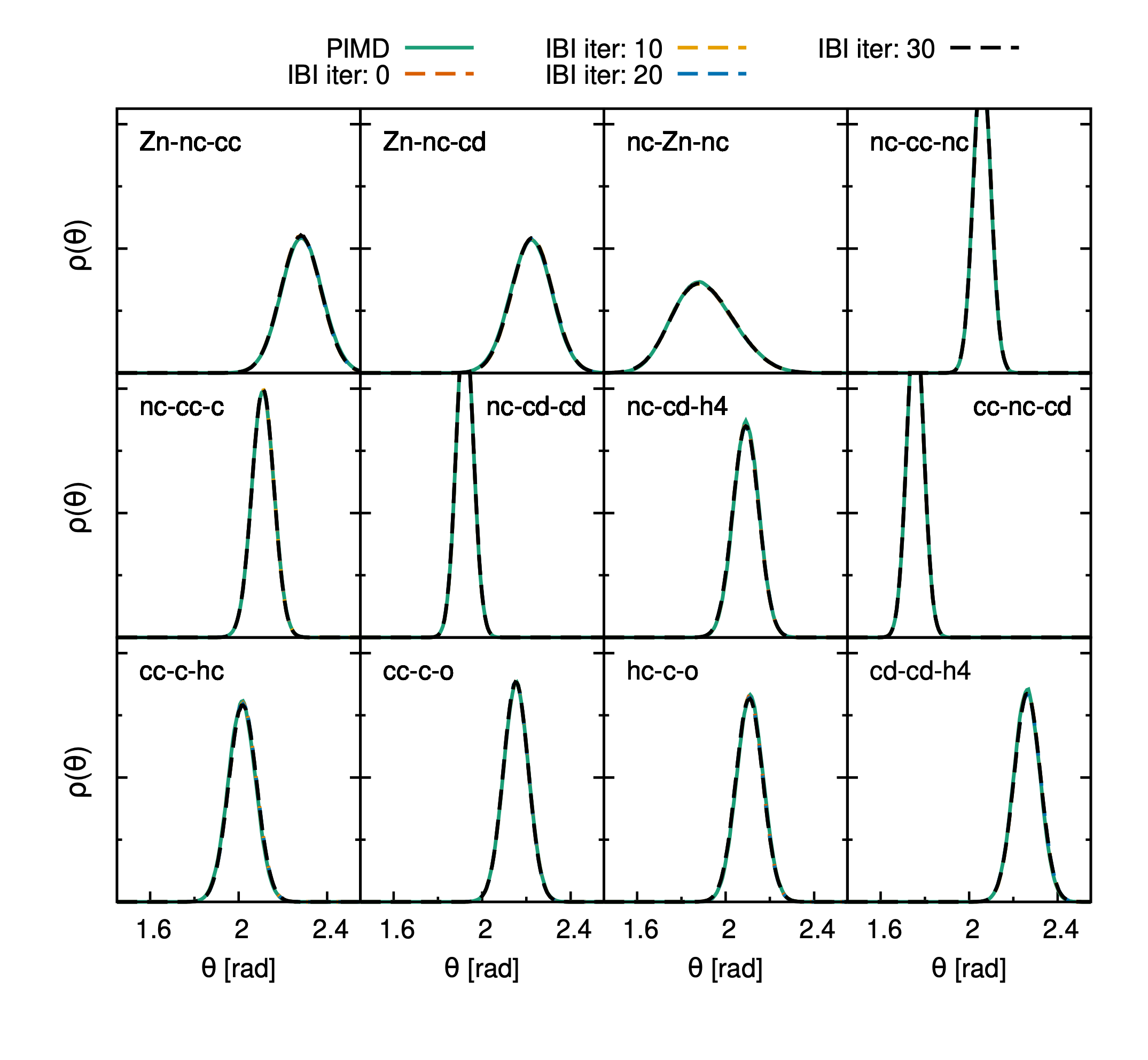} 
\caption{Convergence of the intra-molecular bond angles in ZIF-90 across IBI iterations.}
\label{fig-zif-angle}
\addcontentsline{toc}{subsection}{Figure \ref{fig-zif-angle}. Intra-molecular bond angle distribution of ZIF-90.}
\end{figure}

\clearpage
\begin{figure}[!ht] 
\centering 
\includegraphics[width=0.99\textwidth]{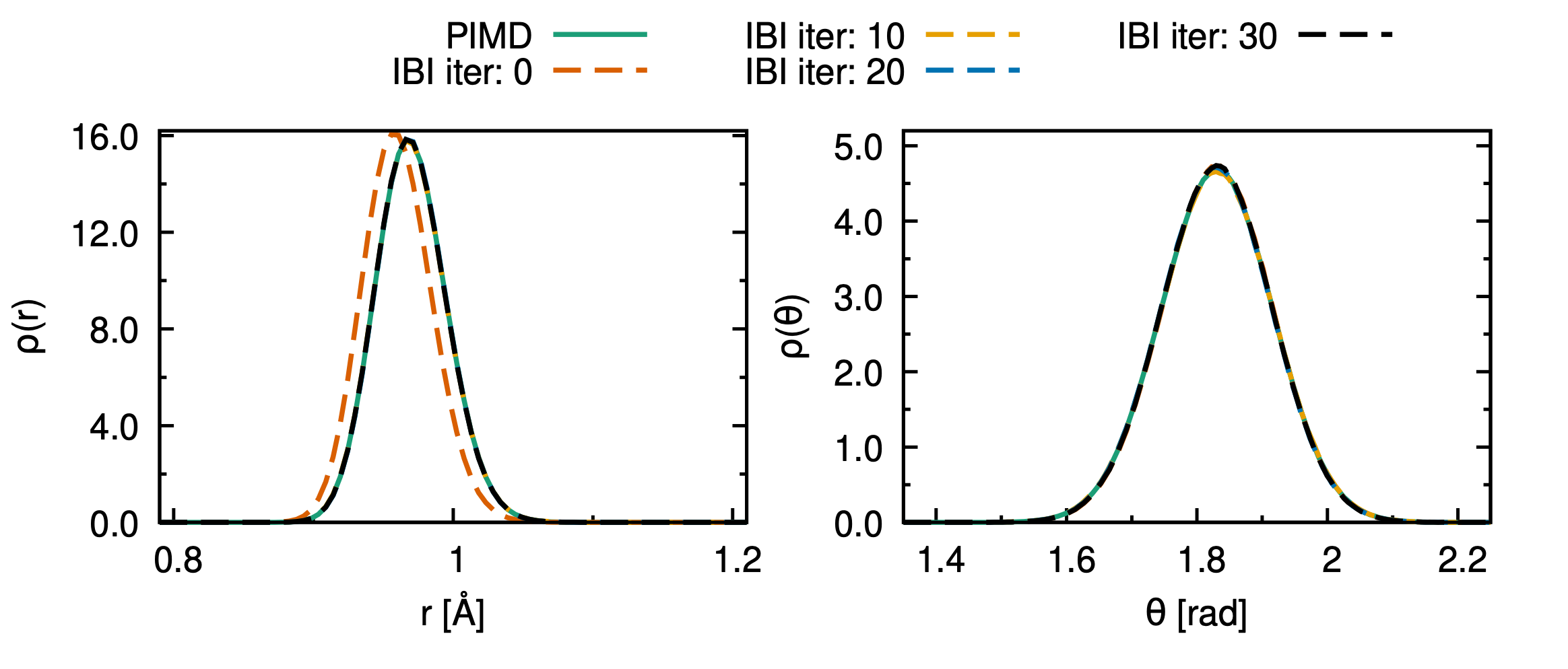} 
\caption{Convergence of the intra-molecular bond lengths and angles of water represented as quasi-centroid distribution functions over IBI iterations. The plot shows the bond lengths and angles approaching the exact PIMD calculation as the number of IBI iterations increases.}
\label{fig-water-qc}
\addcontentsline{toc}{subsection}{Figure \ref{fig-water-qc}. Intra-molecular bond and angle as QC distribution functions of water.}
\end{figure}

\clearpage
\begin{figure}[!ht] 
\centering 
\includegraphics[width=0.99\textwidth]{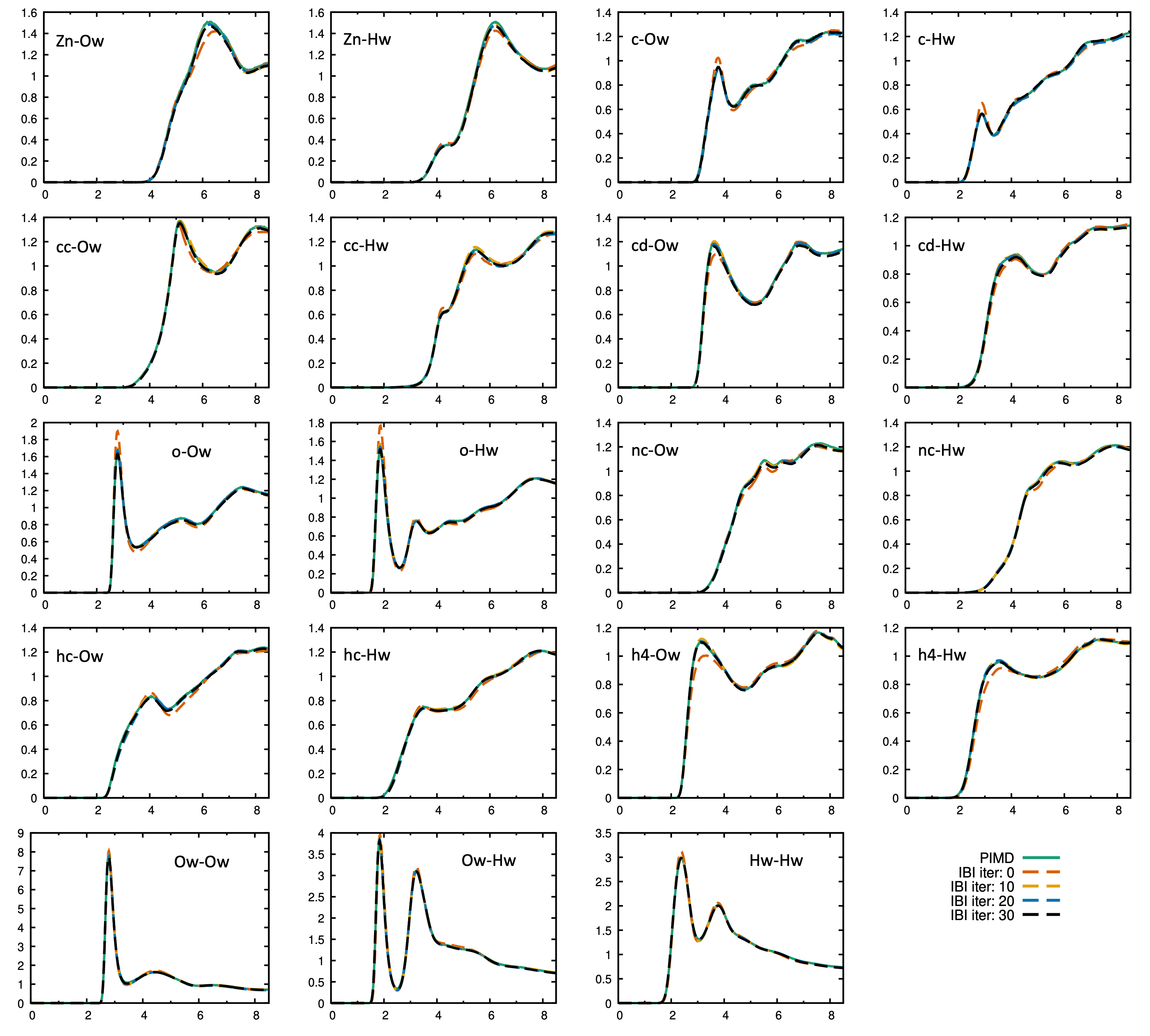} 
\caption{Inter-molecular RDFs of all the pairs of ZIF-90 and water molecules over IBI iterations..}
\label{fig-zif-water-rdf}
\addcontentsline{toc}{subsection}{Figure \ref{fig-zif-water-rdf}. RDFs for all the pairs of ZIF-90 and water system.}
\end{figure}

\clearpage
\begin{figure}[!ht] 
\centering 
\includegraphics[width=0.7\textwidth]{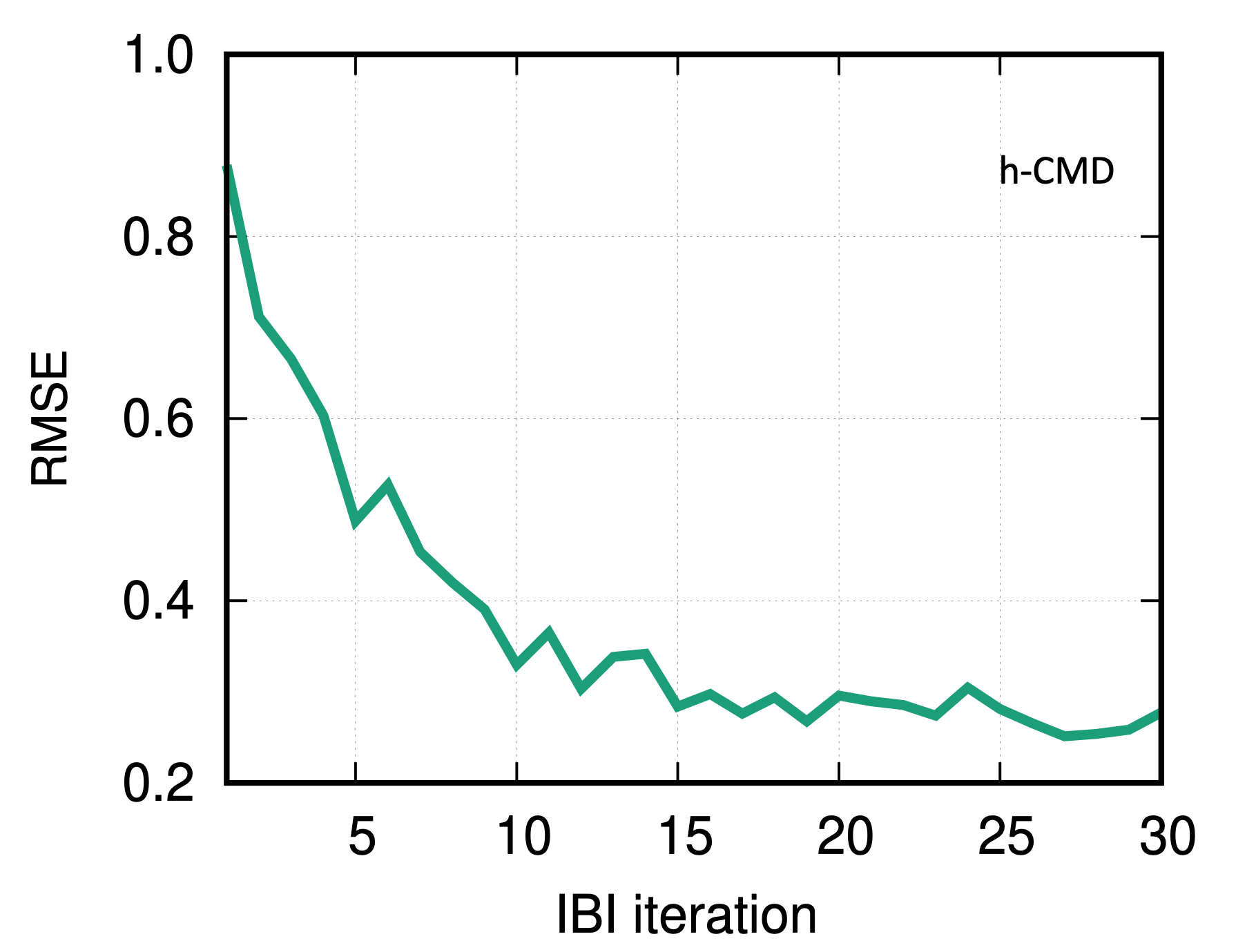} 
\caption{Convergence of the RMSE between reference and computed RDFs across IBI iterations. RDFs for all 19 pairs are used to calculate the RMSE. The results indicate that the correction to the classical potential, $V_{cl}$, converges after 15 IBI iterations for h-CMD.}
\label{fig-rmse}
\addcontentsline{toc}{subsection}{Figure \ref{fig-rmse}. Convergence of the RMSE over IBI iterations.}
\end{figure}
s

\clearpage
\begin{table}
\caption{Average simulation temperatures for ZIF-90 at 60\% RH using different PI methods for all 50 NVE trajectories.}
\addcontentsline{toc}{subsection}{Table \ref{avg_t}. Average simulation temperatures of ZIF-90 at 60\% RH using different PI methods.}
\centering
\setlength{\tabcolsep}{4mm}
\scalebox{0.8}{
\begin{tabular}{c c c c c c}
\midrule[1.4]
Trajectories & MD  & T-RPMD  & PA-CMD & f-CMD & h-CMD  \\
\midrule[1.4]
  1 &  300.3 &  300.1 &  300.2 &  301.5 &  300.5 \\
  2 &  298.9 &  300.2 &  300.1 &  301.2 &  304.6 \\
  3 &  301.8 &  300.1 &  300.1 &  303.4 &  305.1 \\
  4 &  306.4 &  300.0 &  300.1 &  292.6 &  307.9 \\
  5 &  303.7 &  300.1 &  300.2 &  302.2 &  299.8 \\
  6 &  302.7 &  300.1 &  300.1 &  305.2 &  295.1 \\
  7 &  291.5 &  300.2 &  299.8 &  312.5 &  306.6 \\
  8 &  304.0 &  300.2 &  299.9 &  302.8 &  303.2 \\
  9 &  293.8 &  300.2 &  300.0 &  300.5 &  301.1 \\
 10 &  301.5 &  300.1 &  300.2 &  302.1 &  306.9 \\
 11 &  293.7 &  300.1 &  300.1 &  306.9 &  297.6 \\
 12 &  295.7 &  300.3 &  300.1 &  304.4 &  298.9 \\
 13 &  308.8 &  300.1 &  299.8 &  294.7 &  297.1 \\
 14 &  307.9 &  300.1 &  300.3 &  292.7 &  296.9 \\
 15 &  309.6 &  300.1 &  300.1 &  303.2 &  300.9 \\
 16 &  303.9 &  300.1 &  299.9 &  309.4 &  303.4 \\
 17 &  299.9 &  300.0 &  300.0 &  306.8 &  301.7 \\
 18 &  297.6 &  300.0 &  300.1 &  301.7 &  301.5 \\
 19 &  302.6 &  300.0 &  299.9 &  294.9 &  305.4 \\
 20 &  303.5 &  300.1 &  300.1 &  305.4 &  311.2 \\
 21 &  308.9 &  300.3 &  300.2 &  301.8 &  307.2 \\
 22 &  297.1 &  300.2 &  300.7 &  301.9 &  310.6 \\
 23 &  301.6 &  300.1 &  299.9 &  300.6 &  301.6 \\
 24 &  295.6 &  300.1 &  300.3 &  303.5 &  302.7 \\
 25 &  298.2 &  300.2 &  300.4 &  301.7 &  294.5 \\
 26 &  294.4 &  300.1 &  300.2 &  298.8 &  303.1 \\
 27 &  298.9 &  300.2 &  300.1 &  304.6 &  307.0 \\
 28 &  300.2 &  300.2 &  300.1 &  298.3 &  293.8 \\
 29 &  301.4 &  300.1 &  300.2 &  304.9 &  295.3 \\
 30 &  288.4 &  300.2 &  300.3 &  300.7 &  303.1 \\
 31 &  292.9 &  299.9 &  300.0 &  304.1 &  300.4 \\
 32 &  298.1 &  300.1 &  300.5 &  296.8 &  296.9 \\
 33 &  296.4 &  300.3 &  300.0 &  296.5 &  297.6 \\
 34 &  297.7 &  300.1 &  300.2 &  293.6 &  302.0 \\
 35 &  301.4 &  300.3 &  300.0 &  292.3 &  302.7 \\
 36 &  301.8 &  300.1 &  300.3 &  305.9 &  301.0 \\
 37 &  303.7 &  300.1 &  300.5 &  297.0 &  298.7 \\
 38 &  291.5 &  300.2 &  299.9 &  295.1 &  295.4 \\
 39 &  301.0 &  300.2 &  300.4 &  297.3 &  296.0 \\
 40 &  297.0 &  300.1 &  300.2 &  303.7 &  297.1 \\
 41 &  306.9 &  300.1 &  300.2 &  299.1 &  306.8 \\
 42 &  299.3 &  300.1 &  300.0 &  296.9 &  295.5 \\
 43 &  307.2 &  300.1 &  299.8 &  299.2 &  304.2 \\
 44 &  298.2 &  300.1 &  300.0 &  299.8 &  293.3 \\
 45 &  299.6 &  300.1 &  300.1 &  305.4 &  301.0 \\
 46 &  302.2 &  300.1 &  300.1 &  303.9 &  300.7 \\
 47 &  298.0 &  300.1 &  300.2 &  297.1 &  294.8 \\
 48 &  294.8 &  300.1 &  300.4 &  301.0 &  298.6 \\
 49 &  307.5 &  300.2 &  300.1 &  295.5 &  300.3 \\
 50 &  302.7 &  300.2 &  300.4 &  300.9 &  306.5 \\
 \hline
Average & 300.2 & 300.1 & 300.1 & 300.8 & 301.0 \\
\midrule[1.4]
\end{tabular}}     
\label{avg_t}
\end{table}

\clearpage
\bibliography{articles.bib}
\bibliographystyle{achemso}
\addcontentsline{toc}{subsection}{References}